\documentclass[lettersize,hidelinks,journal]{IEEEtran}

\usepackage{cite}
\usepackage{amsmath,amssymb,amsfonts}
\usepackage{graphicx}
\usepackage{textcomp}
\usepackage[dvipsnames]{xcolor}
\usepackage[acronym,shortcuts]{glossaries}
\usepackage{bm}
\usepackage{caption}
\usepackage{subcaption}
\usepackage{relsize}
\usepackage{soul}
\usepackage[bottom]{footmisc}
\usepackage{algorithm, algpseudocode}
\usepackage{algcompatible}
\usepackage{outlines}
\usepackage{orcidlink}
\usepackage{lipsum,comment}
\usepackage{mathtools}
\usepackage{tabularx,flushend}
\usepackage{amsthm}
\usepackage{dblfloatfix}
\newtheorem{theorem}{Theorem}
\theoremstyle{definition}
\newtheorem{lemma}{Lemma}
\newtheorem{definition}{Definition}

\newtheorem{remark}{Remark}

\def\BibTeX{{\rm B\kern-.05em{\sc i\kern-.025em b}\kern-.08em
T\kern-.1667em\lower.7ex\hbox{E}\kern-.125emX}}

\newcommand{\trans}[0]{^{\mathsf{T}}}

\newcommand{\herm}[0]{^{\mathsf{H}}}

\newcommand{\inner}[2]{\langle #1, #2 \rangle}

\newacronym{RPE}{RPE}{radar parameter estimation}
\newacronym{OTFS}{OTFS}{orthogonal time frequency space}
\newacronym{AFDM}{AFDM}{affine frequency division multiplexing}
\newacronym{MIMO}{MIMO}{multiple-input multiple-output}
\newacronym{SISO}{SISO}{single-input single-output}
\newacronym{ISAC}{ISAC}{integrated sensing and communications}
\newacronym{3D}{3D}{three-dimensional}
\newacronym{2D}{2D}{two-dimensional}
\newacronym{1D}{1D}{one-dimensional}
\newacronym{RX}{RX}{receiver}
\newacronym{TX}{TX}{transmitter}
\newacronym{BF}{BF}{beamforming}
\newacronym{mmWave}{mmWave}{millimeter-wave}
\newacronym{SotA}{SotA}{state-of-the-art}
\newacronym{ULA}{ULA}{uniform linear array}
\newacronym{QAM}{QAM}{quadrature amplitude modulation}
\newacronym{ISFFT}{ISFFT}{inverse symplectic finite Fourier transform}
\newacronym{SFFT}{SFFT}{symplectic finite Fourier transform}
\newacronym{AWGN}{AWGN}{additive white Gaussian noise}
\newacronym{OFDM}{OFDM}{orthogonal frequency division multiplexing}
\newacronym{OCDM}{OCDM}{orthogonal chirp division multiplexing}
\newacronym{BS}{BS}{base station}
\newacronym{UE}{UE}{user equipment}
\newacronym{DFT}{DFT}{discrete Fourier transform}
\newacronym{IDFT}{IDFT}{inverse discrete Fourier transform}
\newacronym{TD}{TD}{time-domain}
\newacronym{wlg}{wlg}{without loss of generality}
\newacronym{CP}{CP}{cyclic prefix}
\newacronym{DAFT}{DAFT}{discrete affine Fourier transform}
\newacronym{IDAFT}{IDAFT}{inverse discrete affine Fourier transform}
\newacronym{CPP}{CPP}{\textit{chirp-periodic} prefix}
\newacronym{IDZT}{IDZT}{inverse discrete Zak transform}
\newacronym{DZT}{DZT}{discrete Zak transform}
\newacronym{ICI}{ICI}{inter-carrier interference}
\newacronym{BER}{BER}{bit error rate}
\newacronym{DoF}{DoF}{degrees-of-freedom}
\newacronym{FD}{FD}{full-duplex}
\newacronym{SIMO}{SIMO}{single-input multiple-output}
\newacronym{MISO}{MISO}{multiple-input single-output}
\newacronym{AoD}{AoD}{angle-of-departure}
\newacronym{AoA}{AoA}{angle-of-arrival}
\newacronym{RF}{RF}{radio frequency}
\newacronym{SIM}{SIM}{stacked intelligent metasurfaces}
\newacronym{FIM}{FIM}{flexible intelligent metasurface}
\newacronym{FPGA}{FPGA}{field programmable gate array}
\newacronym{UPA}{UPA}{uniform planar array}
\newacronym{CC}{CC}{communication-centric}
\newacronym{I/O}{I/O}{input-output}
\newacronym{iid}{i.i.d.}{independent and identically distributed}
\newacronym{IoT}{IoT}{internet of things}
\newacronym{V2X}{V2X}{vehicle-to-everything}
\newacronym{NTN}{NTN}{non-terrestrial network}
\newacronym{LEO}{LEO}{low-earth orbit}
\newacronym{THz}{THz}{terahertz}
\newacronym{EM}{EM}{electromagnetic}
\newacronym{RIS}{RIS}{reconfigurable intelligent surface}
\newacronym{DoA}{DoA}{direction-of-arrival}
\newacronym{DD}{DD}{doubly-dispersive}
\newacronym{ODDM}{ODDM}{orthogonal delay-Doppler division multiplexing}
\newacronym{LoS}{LoS}{line-of-sight}
\newacronym{NLoS}{NLoS}{non-line-of-sight}
\newacronym{6G}{6G}{sixth generation}
\newacronym{MPDD}{MPDD}{metasurfaces-parameterized DD}
\newacronym{GaBP}{GaBP}{Gaussian Belief Propagation}
\newacronym{MSE}{MSE}{mean-squared-error}
\newacronym{sIC}{soft IC}{soft interference cancellation}
\newacronym{soft RG}{soft RG}{soft replica generation}
\newacronym{BG}{BG}{belief generation}
\newacronym{SGA}{SGA}{scalar Gaussian approximation}
\newacronym{CLT}{CLT}{central limit theorem}
\newacronym{PDF}{PDF}{probability density function}
\newacronym{QPSK}{QPSK}{quadrature phase-shift keying}
\newacronym{LMMSE}{LMMSE}{linear minimum mean square error}
\newacronym{SNR}{SNR}{signal-to-noise ratio}
\newacronym{QoS}{QoS}{quality of service}
\newacronym{CoV}{CoV}{calculus of variations}
\newacronym{CAPA}{CAPA}{continuous aperture array}
\newacronym{GL}{GL}{Gauss-Legendre}
\newacronym{DDC MIMO}{DDC MIMO}{DD continuous MIMO}
\newacronym{B5G}{B5G}{beyond fifth generation}
\newacronym{VR}{VR}{virtual reality}
\newacronym{XR}{XR}{extended reality}
\newacronym{ITN}{ITN}{intelligent traffic networks}
\newacronym{SAGIN}{SAGIN}{space-air-ground integrated network}
\newacronym{UAV}{UAV}{unmanned aerial vehicle}
\newacronym{MUSIC}{MUSIC}{Multiple Signal Classification}
\newacronym{ICC}{ICC}{integrated communication and computing}
\newacronym{MRC}{MRC}{maximum ratio combining}

\begin{document}

% \title{Channel Modeling and Parametrization of Doubly-Dispersive Continuous MIMO Systems}
%\title{Doubly-Dispersive Continuous MIMO Systems: \\ Channel Modeling and Beamforming Optimization}

\title{Doubly-Dispersive Continuous MIMO Systems: \\ Channel Modeling and Beamforming Design}

% \title{\color{blue}Continuous MIMO in Doubly-Dispersive Channels: \\ Fundamental Modeling and Beamforming Design} % just a suggestion

\author{Kuranage Roche Rayan Ranasinghe\textsuperscript{\orcidlink{0000-0002-6834-8877}},~\IEEEmembership{Graduate Student Member,~IEEE,}\,
Zhaolin Wang\textsuperscript{\orcidlink{0000-0003-4614-0175}},~\IEEEmembership{Member,~IEEE,} \\ 
Hyeon Seok Rou\textsuperscript{\orcidlink{0000-0003-3483-7629}},~\IEEEmembership{Member,~IEEE,} Giuseppe Thadeu Freitas de Abreu\textsuperscript{\orcidlink{0000-0002-5018-8174}},~\IEEEmembership{Senior Member,~IEEE,} \\ and Emil Bj{\"o}rnson\textsuperscript{\orcidlink{0000-0002-5954-434X}},~\IEEEmembership{Fellow,~IEEE}
% <-this % stops a space 
\thanks{K. R. R. Ranasinghe, H. S. Rou, and G. T. F. de Abreu are with the School of Computer Science and Engineering, Constructor University (previously Jacobs University Bremen), Campus Ring 1, 28759 Bremen, Germany (emails: \{kranasinghe, hrou, gabreu\}@constructor.university).}
\thanks{Z. Wang is with the Department of Electrical and Electronic Engineering, The University of Hong Kong, Hong Kong (email: zhaolin.wang@hku.hk).}
\thanks{E. Bj{\"o}rnson is with the School of Electrical Engineering and Computer Science, KTH Royal Institute of Technology, Stockholm 16440, Sweden (email: emilbjo@kth.se).}
\vspace{-3ex}}

%\markboth{To be submitted to the IEEE for possible publication}{Shell \MakeLowercase{\textit{et al.}}: A Sample Article Using IEEEtran.cls for IEEE Journals}


\maketitle

\begin{abstract}
We address the modeling and optimal \ac{BF} design for \ac{MIMO} \ac{CAPA} systems operating over \ac{DD} channels.
First, a comprehensive \ac{DDC MIMO} channel model that incorporates \acp{CAPA} at both the \ac{TX} and \ac{RX} is derived, which is used to obtain explicit \ac{I/O} relations for various waveforms well suited to \ac{ISAC} and robust to \ac{DD} channels, namely \ac{OFDM}, \ac{OTFS}, and \ac{AFDM}.
Then, functional optimization problems are formulated for the design of \ac{TX} and \ac{RX} \ac{BF} matrices that maximize received power, in which novel low-complexity, closed-form solutions are obtained via the \ac{CoV} method, yielding expressions closely related to the classical matched filter commonly used in conventional \ac{MIMO} systems.
Simulation results confirm that the proposed \ac{TX}/\ac{RX} \ac{BF} designs with \acp{CAPA} provide significant performance and computational complexity gains over conventional \ac{MIMO} systems in \ac{DD} channels.
\end{abstract}


\glsresetall

\vspace{-1ex}
\section{Introduction}

%% Intro. to doubly-dispersive channels 
%% Into to 6G , ....

\IEEEPARstart{C}{urrent} research on upcoming \ac{B5G} and \ac{6G} wireless networks is increasingly driven by extraordinary performance demands and the need to support a wide range of demanding applications, including \ac{ISAC} \cite{LuongCOMMST2025}, \ac{VR}/\ac{XR} \cite{YuNWTWORK2023}, massive \ac{IoT} \cite{Chowdhury_6G}, \ac{ICC} \cite{ranasinghe2025flexibledesignframeworkintegrated}, \ac{SAGIN} \cite{CuiCC2022}, holographic communications \cite{DengJSAC2023}, and \ac{ITN} \cite{NguyenJSAC2024}.
Meeting these diverse requirements and applications will require \ac{6G} systems to operate in challenging wireless environments, especially in high-mobility multipath scenarios such as \ac{LEO} satellites, \ac{V2X}, \ac{UAV}, and high-speed railways. 
These conditions give rise to the so-called \ac{DD} channel \cite{Bliss_Govindasamy_2013}, where Doppler-domain spreading introduces severe impairments to conventional \ac{OFDM} \cite{wang2006performance}. This has motivated the development of next-generation waveforms \cite{RouSPM2024}, including \ac{OTFS} \cite{Hadani_WCNC_2017, wei2021orthogonal} and \ac{AFDM} \cite{Bemani_TWC_2023,rou2025affine}.
By leveraging the full delay-Doppler representation of the \ac{DD} channel, these waveforms have also been shown to result in beneficial implications to enable native \emph{communication-centric} \ac{ISAC} techniques \cite{RanasingheTWC2025,yuan2024otfs}.

%% EM strcutures
In parallel with the above, an alternative approach to achieving unprecedented network capacity, data rate per device, and sensing accuracy is to embed a large number of antennas within a relatively small surface area, leading to the emerging extra-large (XL) -- and to the \textit{gigantic} -- \ac{MIMO} paradigm \cite{BjornsonOJCOMS2025}.
To this end, several \ac{SotA} array architectures have recently been explored, including \ac{RIS} \cite{DiRenzoJSAC2020,ZhangTWC2025}, \acp{SIM} \cite{AnTWC2025,RanasingheARXIV2025}, and, most recently, \acp{FIM} \cite{AnTWC2025FIM,ranasinghe2025FIM}.
Following the above, this trend toward increasingly larger, denser, low-cost, and flexible \ac{EM} structures, enabled by emerging technologies based on advanced materials, has recently motivated the concept of \acp{CAPA}.

Owing to their inherent continuous nature, the modeling and optimization of \acp{CAPA} rely on fundamental \ac{EM} theories, which map the continuous radiating surface into information-bearing sinusoidal source currents distributed across the transmit aperture. These currents generate \ac{EM} waves that are subsequently received and decoded by a receive \ac{CAPA} to recover the desired information.
While extensive literature exists on \ac{BF} design for \ac{CAPA} systems (see, e.g., \cite{WangTWC2025,wang2025beamformingdesigncontinuousaperture,OuyangTWC2025}), channel modeling for such systems remains limited (e.g., \cite{OuyangTWC2025Diversity}), with no prior work addressing the \ac{DD} characteristics typical of high-mobility channels as expected in the imminent generations.

Therefore, we address this gap by proposing a novel \ac{DDC MIMO} channel model that captures the practical \ac{DD} nature of wireless channels between two \acp{CAPA}, derived from the fundamental scattering matrix model and dyadic Green's function.
Furthermore, based on the derived model, we obtain explicit \ac{I/O} relations for the \ac{OFDM}, \ac{OTFS}, and \ac{AFDM} waveforms, where the latter have been extensively shown to perform effectively in \ac{DD} channels.
Finally, we formulate and solve a receive power maximization problem for the proposed \ac{DDC MIMO} system, and obtain closed-form optimal continuous beamformers for both \ac{TX} and \ac{RX} using the \ac{CoV} technique.

% In all, the contributions of this paper are summarized as follows:
% %
% \begin{itemize}
% %
% \item A novel \ac{DDC MIMO} channel model that incorporates \ac{MIMO} \acp{CAPA} at both the \ac{TX} and \ac{RX}.
% %
% \item We derive the explicit \ac{I/O} relationships for \ac{OFDM}, \ac{OTFS}, and \ac{AFDM} waveforms under the proposed \ac{DDC MIMO} model.
% %
% \end{itemize}

% \begin{itemize}
% \item Finally, a functional optimization problem to maximize the received power is  formulated and solved via the \ac{CoV} technique, yielding closed-form \ac{BF} designs for both the transmitter and receiver.
% %
% \end{itemize}

\textit{Notation:} Scalars are denoted by uppercase or lowercase letters, column vectors by bold lowercase letters, and matrices by bold uppercase letters.
The diagonal matrix constructed from vector $\mathbf{a}$ is denoted by $\text{diag}(\mathbf{a})$. For a matrix $\mathbf{A}$, we use $\mathbf{A}\trans$, $\mathbf{A}\herm$, $\mathbf{A}^{1/2}$, and $[\mathbf{A}]_{i,j}$ to represent its transpose, Hermitian, square root, and $(i,j)$-th element, respectively.
The convolution and Kronecker product are denoted by $*$ and $\otimes$, while $\mathbf{I}_N$ and $\mathbf{F}_N$ denote the $N \times N$ identity matrix and the normalized $N$-point \ac{DFT} matrix, respectively.
The sinc function is defined as $\text{sinc}(a) \triangleq \frac{\sin(\pi a)}{\pi a}$, and $j \triangleq \sqrt{-1}$ denotes the imaginary unit.
The Dirac delta function is denoted by $\delta(\cdot)$.
The Lebesgue measure of a Euclidean subspace $\mathcal{S}$ is denoted by $|\mathcal{S}|$.
The absolute value and Euclidean norm are denoted by $|\cdot|$ and $\|\cdot\|$, respectively.

% \subsection{Continuous Channel with Doppler Frequencies}

%%%%%%%%%%%%%%%%%%%%%%%%%%%%%%%%%%%%%%%%%%%%%%%%%%%%%
\section{Continuous DD MIMO Channel Model}
\label{FIM_MIMO_Model}

\begin{figure}[b!]
\vspace{-3ex}
\centering
\includegraphics[width=0.93\columnwidth]{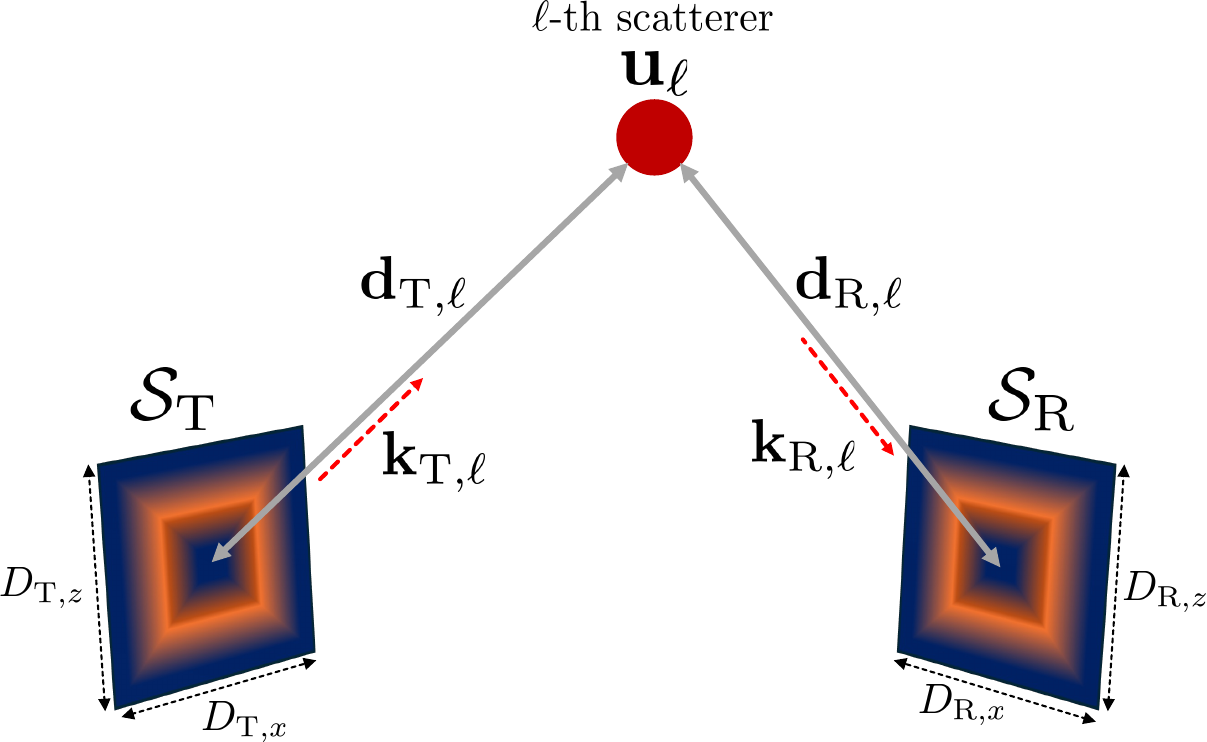}
\caption{Illustration of the point-to-point \ac{DDC MIMO} system.}
\label{fig:system_model}
\vspace{-1.5ex}
\end{figure}

% \Rayan{If we really keep the array structures as general as possible till the end, we could even show how we can derive continuous versions of ALL the artificial structures like SIMS/FIMS and CAPAs!!}
Consider a point-to-point \ac{MIMO} system with \acp{CAPA} at both the \ac{TX} and \ac{RX}. The transmit \ac{CAPA} occupies a continuous surface $\mathcal{S}_{\mathrm{T}}$ of area $A_\mathrm{T} = |\mathcal{S}_{\mathrm{T}}|$, while the receive \ac{CAPA} occupies a continuous surface $\mathcal{S}_{\mathrm{R}}$ of area $A_\mathrm{R} = |\mathcal{S}_{\mathrm{R}}|$, as illustrated in Fig.~\ref{fig:system_model}.
\Ac{wlg}, the square-shaped transmit \ac{CAPA} lies on the $x\text{-}z$ plane, centered at the origin. Its two sides are parallel to the $x$- and $z$-axes, with side lengths denoted by $D_{\mathrm{T},x}$ and $D_{\mathrm{T},z}$, respectively.
An arbitrary time-domain transmit signal radiated from a point $\mathbf{s} \!=\! [s_x,0,s_z]\trans \!\in \mathcal{S}_\mathrm{T}$ at time $t$ is represented as $\bm{x}(\mathbf{s},t) \!\in \mathbb{C}^{3 \times 1}$, with its frequency-domain counterpart defined by
\begin{equation}
\label{eq:Freq-domain-transmit}
\bm{x}(\mathbf{s}, \omega) = \int_{-\infty}^{+\infty} \bm{x}(\mathbf{s},t) e^{-j \omega t} d t \in \mathbb{C}^{3 \times 1}, 
\end{equation}
where $\omega = 2\pi f$ denotes the angular frequency, with $f$ being the signal frequency.

Similarly, an arbitrary time-domain receive signal impinging on a point $\mathbf{r} = [r_x,D_\mathrm{RT},r_z]\trans \in \mathcal{S}_\mathrm{R}$ of the square-shaped receive \ac{CAPA}\footnote{$D_\mathrm{RT}$ denotes the physical $y$-axis separation between the transmit and receive \acp{CAPA}.
Extensions to arbitrary coordinate systems are straightforward and can be carried out in the same manner as in \cite{wang2025beamformingdesigncontinuousaperture} without affecting the fundamental results presented in this paper.}, which lies on the $x\text{-}z$ plane with side lengths $D_{\mathrm{R},x}$ and $D_{\mathrm{R},z}$ along the $x$- and $z$-axes, is represented at time $t$ as $\bm{y}(\mathbf{r},t) \in \mathbb{C}^{3 \times 1}$. Its frequency-domain counterpart is defined as
\begin{equation}
\label{eq:Freq-domain-receive}
\bm{y}(\mathbf{r}, \omega) = \int_{-\infty}^{+\infty} \bm{y}(\mathbf{r},t) e^{-j \omega t} \, dt \in \mathbb{C}^{3 \times 1}. 
\end{equation}

Next, following \cite{PoonTIT2005}, the transfer function between the transmitted and received signals in the frequency domain, defined in equations \eqref{eq:Freq-domain-transmit} and \eqref{eq:Freq-domain-receive}, respectively, can be expressed as
\vspace{-1ex}
\begin{equation}
\label{eq:Freq-domain-transfer_function}
\bm{y}(\mathbf{r}, \omega) = \int_{\mathcal{S}_\mathrm{T}} \bm{H}(\mathbf{r},\mathbf{s}) \bm{x}(\mathbf{s}, \omega) d\mathbf{s} + \bm{z}(\mathbf{r}) \in \mathbb{C}^{3 \times 1},
\end{equation}
where the term $\bm{z}(\mathbf{r}) \in \mathbb{C}^{3 \times 1}$ describes the \ac{AWGN}.

For the aforementioned continuous \ac{MIMO} scenario, the channel response $\bm{H}(\mathbf{r},\mathbf{s}) \in \mathbb{C}^{3 \times 3}$ seen in equation \eqref{eq:Freq-domain-transfer_function} which incorporates both the channel gain and polarization between the transmit and receive positions $\mathbf{s}$ and $\mathbf{r}$, respectively, can be modeled as
\begin{equation}
\label{eq:Freq-domain-channel_response}
\!\!\bm{H}(\mathbf{r},\mathbf{s}) \!=\!\! \int_{\mathbf{\Omega}} \! \int_{\mathbf{\Omega}} \!\!\bm{G}_{\mathrm{R}} (\mathbf{r}, \mathbf{q}) \bm{C}(\mathbf{q}, \mathbf{p}) \bm{G}_{\mathrm{T}} (\mathbf{p}, \mathbf{s}) d\mathbf{p} d\mathbf{q},\!\!
\end{equation}
where the double integral over $\mathbf{\Omega}$ defines the entirety of the \ac{3D} space with $\mathbf{p}$ and $\mathbf{q}$ respectively denoting arbitrary transmit and receive directions from the \ac{TX} to the scatterers and from the scatterers to the \ac{RX}.

In equation \eqref{eq:Freq-domain-channel_response}, $\bm{G}_{\mathrm{T}} (\mathbf{p}, \mathbf{s}) \in \mathbb{C}^{3 \times 3}$ and $\bm{G}_{\mathrm{R}} (\mathbf{r}, \mathbf{q}) \in \mathbb{C}^{3 \times 3}$ respectively denote the transmit and receive array responses which map the current distributions on the \ac{TX} and the \ac{RX} to the radiated/incident field patterns, while $\bm{C}(\mathbf{q}, \mathbf{p}) \in \mathbb{C}^{3 \times 3}$ is the response due to the $L$ scatterers including the channel gain and polarization effects, between the transmit and receive directions $\mathbf{p}$ and $\mathbf{q}$, respectively.

\renewcommand{\arraystretch}{1.25}
\setlength{\tabcolsep}{1.2pt}
\begin{table}[t]
\caption{Table of Symbols for Variables and Parameters}
\centering
\begin{tabular}{|c|c|}
\hline
\textbf{Variable} & \textbf{Description} \\
\hline
$\mathcal{S}_{\mathrm{T}}$, $\mathcal{S}_{\mathrm{R}}$ & Euclidean subspaces of the TX and RX \\
\hline
$\mathbf{s}_{t_0}$, $\mathbf{s}_{r_0}$ & Centers of $\mathcal{S}_{\mathrm{T}}$ and $\mathcal{S}_{\mathrm{R}}$ \\
\hline
$\mathbf{s}$, $\mathbf{r}$ & Arbitrary points on $\mathcal{S}_{\mathrm{T}}$ and $\mathcal{S}_{\mathrm{R}}$ \\
\hline
$L$ & Number of scatterers \\
\hline
$\mathbf{u}_\ell$ & Position vector of the $\ell$-th scatterer \\
\hline
$\mathbf{v}_{\mathrm{T},\ell}$ & Velocity vector of the $\ell$-th scatterer \\
\hline
$\mathbf{d}_{\mathrm{T},\ell}$, $\mathbf{d}_{\mathrm{R},\ell}$ & Distance of the $\ell$-th scatterer from the TX and RX \\ 
\hline
$\mathbf{k}_{\mathrm{T},\ell}$, $\mathbf{k}_{\mathrm{R},\ell}$ & Direction of the $\ell$-th scatterer from the TX and RX \\
\hline
$\mathbf{p}$, $\mathbf{q}$ & Arbitrary direction vectors from the TX and RX \\
\hline
$\tau_\ell$ & Total path delay of the $\ell$-th scatterer between the TX and RX\\
\hline
$\nu_\ell$ & Total bistatic Doppler shift of the $\ell$-th scatterer\\
\hline
\end{tabular}
\label{tab:example}
\vspace{-2ex}
\end{table}

\vspace{-2ex}
\subsection{Scattering Matrix Response Model}

Following classical physics-based ray-tracing models \cite{ZwickJSAC2002,PoonTIT2005} as well as more recent \ac{DD} models \cite{RouSPM2024,RanasingheTWC2025} used in discrete \ac{MIMO} \cite{RanasingheARXIV2025} settings, the scattering matrix response can be modeled as
\vspace{-1ex}
\begin{equation}
% \vspace{-1ex}
\label{eq:scattering_matrix_resp}
\vspace{-1ex}
\bm{C}(\mathbf{q}, \mathbf{p}) = \frac{1}{\sqrt{L}} \sum_{\ell=1}^L \mathbf{\Gamma}_\ell \delta(\mathbf{q} - \mathbf{u}_\ell) \delta(\mathbf{p} - \mathbf{u}_\ell),
\end{equation}
where $L$ denotes the total number of scatterers\footnote{The factor $1/\sqrt{L}$  normalizes the power of the scattered signal \cite{PoonTIT2005}.}, $\mathbf{u}_\ell$ denotes the position vector of the $\ell$-th scatterer and $\mathbf{\Gamma}_\ell \in \mathbb{C}^{3 \times 3}$ comprises all attenuation and polarization effects caused by a given scatterer.

It is noteworthy that the aforementioned model is general in the sense that it encompasses both the far field and near field scenarios by modeling a scatterer as an impulse response in free space.
Leveraging the properties of the Dirac delta function, substituting equation \eqref{eq:scattering_matrix_resp} in equation \eqref{eq:Freq-domain-channel_response} yields \vspace{-1ex}
\begin{align}
\vspace{-1ex}
\label{eq:Freq-domain-channel_response_w_scattering_matrix}
\bm{H}(\mathbf{r},\mathbf{s}) &= \frac{1}{\sqrt{L}} \sum_{\ell=1}^L \int_{\mathbf{\Omega}} \! \int_{\mathbf{\Omega}} \!\!\bm{G}_{\mathrm{R}} (\mathbf{r}, \mathbf{q}) \delta(\mathbf{q} - \mathbf{u}_\ell) \mathbf{\Gamma}_\ell \delta(\mathbf{p} - \mathbf{u}_\ell) \nonumber \\
&\hspace{12ex}\times \bm{G}_{\mathrm{T}} (\mathbf{p}, \mathbf{s}) d\mathbf{p} d\mathbf{q} \nonumber \\
&= \frac{1}{\sqrt{L}} \sum_{\ell=1}^L \bigg( \int_{\mathbf{\Omega}} \!\!\bm{G}_{\mathrm{R}} (\mathbf{r}, \mathbf{q}) \delta(\mathbf{q} - \mathbf{u}_\ell) d\mathbf{q} \bigg) \mathbf{\Gamma}_\ell \nonumber \\
&\hspace{12ex}\times \bigg( \int_{\mathbf{\Omega}} \!\!\bm{G}_{\mathrm{T}} (\mathbf{p}, \mathbf{s}) \delta(\mathbf{p} - \mathbf{u}_\ell) d\mathbf{p} \bigg) \nonumber \\[-1ex]
&= \frac{1}{\sqrt{L}} \sum_{\ell=1}^L \bm{G}_{\mathrm{R}} (\mathbf{r}, \mathbf{u}_\ell) \mathbf{\Gamma}_\ell \bm{G}_{\mathrm{T}} (\mathbf{u}_\ell, \mathbf{s}),
\end{align}
\vspace{-2ex}

\noindent where we have used the facts that $\delta(\mathbf{a} - \mathbf{b}) = 0, \forall \mathbf{a} \neq \mathbf{b}$, $\int_{\mathbf{\Omega}} \delta(\mathbf{a} - \mathbf{b}) d\mathbf{a} = 1$ and $\int_{\mathbf{\Omega}} \bm{G}(\mathbf{a}) \delta(\mathbf{a} - \mathbf{b}) d\mathbf{a} = \bm{G}(\mathbf{b})$, with $\mathbf{a}, \mathbf{b} \in \mathbb{C}^{3 \times 1}$ denoting arbitrary vectors and $\bm{G}(\mathbf{a})$ denoting an arbitrary continuous array response.

\vspace{-1ex}
\subsection{Continuous Array Response Model}

It is well established \cite{PoonTIT2005,ZwickJSAC2002,wei2024electromagneticinformationtheoryholographic,YangTWC2024,WeiTWC2025} that continuous array responses, which map current distributions on the arrays to the resulting \ac{EM} field patterns, can be described using the dyadic Green's function, expressed as \cite{9650519}
\begin{equation}
\label{eq:Greens_function}
\bm{G}(\mathbf{a}, \mathbf{b}) = \left( \mathbf{I}_3 + \frac{\nabla \nabla}{\kappa^2} \right) g(\mathbf{a}, \mathbf{b}),
\end{equation}
where $\kappa = \omega/c$ is the wavenumber, $\nabla$ is the vector differential operator in the three-dimensional Cartesian coordinate system, and $g(\mathbf{a}, \mathbf{b})$ is the scalar Green's function, given by 
\begin{equation}
g(\mathbf{a}, \mathbf{b}) = \frac{1}{4\pi\|\mathbf{a} - \mathbf{b}\|} e^{-j \kappa \|\mathbf{a} - \mathbf{b}\|}.
\end{equation}

For a mobile scatterer inducing a Doppler shift, we have
\begin{equation}
\label{eq:moving_scaterer}
\mathbf{u}_\ell = \mathbf{u}_{\ell,0} + \mathbf{v}_{\mathrm{T},\ell}\cdot t,
\end{equation}
where $\mathbf{u}_{\ell,0} \in \mathbb{C}^{3 \times 1}$ and $\mathbf{v}_{\mathrm{T},\ell} \in \mathbb{C}^{3 \times 1}$ are the initial position and velocity of a $\ell$-th scatterer, respectively.

\vspace{-1ex}
\subsection{Far-Field Approximations}

To derive tractable expressions for the effective channel model, we adopt standard far-field assumptions. 
For clarity, the derivation is presented on the \ac{TX} side, where by symmetry, identical forms hold on the \ac{RX} side. 

Let $\mathbf{u}_\ell$ denote the location of the $\ell$-th scatterer, $\mathbf{s}$ a point on the transmit aperture, $\mathbf{d}_{\mathrm{T},\ell}$ the vector from the TX reference to $\mathbf{u}_\ell$, and $\mathbf{k}_{\mathrm{T},\ell}$ the unit departure direction vector. 
Under the far-field regime, the following approximations hold \cite{10934790,richards2005fundamentals,9650519}.
% \begin{align}
%     &\text{(Far-field phase approximation: )} \nonumber \\
%     & \|\mathbf{u}_\ell - \mathbf{s}\| \approx \underbrace{\|\mathbf{d}_{\mathrm{T},\ell}\| - \mathbf{k}_{\mathrm{T}, \ell}\trans  \mathbf{s}}_{\text{plane-wave linear phase}} \hspace{3ex}+ \underbrace{\mathbf{k}_{\mathrm{T},\ell}\trans \mathbf{v}_{\mathrm{T},\ell} t}_{\text{radial velocity projection}}, \\
%     &\text{(Aperture-amplitude variation neglected: )}\nonumber \\
%     & 1/\|\mathbf{u}_\ell - \mathbf{s}\| \approx 1/\| \mathbf{d}_{\mathrm{T},\ell}\|, \\
%     &\text{(Dyadic operator for a plane wave: )} \nonumber \\
%     &\nabla \nabla \approx - \kappa^2 \mathbf{k}_{\mathrm{T},\ell} \mathbf{k}_{\mathrm{T},\ell}\trans.
% \end{align}

\vspace{0.5ex}
\noindent \textbf{Plane-wave Phase Approximation:} This approximation is the first-order Taylor expansion of the path length, which separates into a constant range term $d_{\mathrm{T},\ell} \triangleq \|\mathbf{d}_{\mathrm{T},\ell}\|$, a linear phase term $\mathbf{k}_{\mathrm{T}, \ell}\trans\mathbf{s}$ across the aperture that represents a plane wave, and a time-varying term caused by the radial velocity component $\tilde{v}_{\mathrm{T},\ell} \triangleq \mathbf{k}_{\mathrm{T},\ell}\trans\mathbf{v}_{\mathrm{T},\ell}$ that produces a Doppler shift, implying
\vspace{-1ex}
\begin{equation}
\|\mathbf{u}_\ell - \mathbf{s}\| \approx d_{\mathrm{T},\ell} - \mathbf{k}_{\mathrm{T}, \ell}\trans\mathbf{s} + \tilde{v}_{\mathrm{T},\ell}t.
\vspace{-1ex}
\end{equation}
\noindent\textbf{Negligible Aperture-Amplitude Variation:} The amplitude variation due to spherical spreading is assumed to be negligible across the aperture and can thus be treated as constant w.r.t. to the center of the surface, yielding
\vspace{-1ex}
\begin{equation}
\vspace{-1ex}
\frac{1}{\|\mathbf{u}_\ell - \mathbf{s}\|} \approx \frac{1}{d_{\mathrm{T},\ell}}.
\end{equation}

\noindent\textbf{Plane-Wave Dyadic Operator:} Using the plane-wave identity $\nabla e^{j\kappa\,\mathbf{k}_{\mathrm{T},\ell}\trans\mathbf{s}} = j\kappa\,\mathbf{k}_{\mathrm{T},\ell} e^{j\kappa\,\mathbf{k}_{\mathrm{T},\ell}\trans\mathbf{s}}$, we approximate
\vspace{-1ex}
\begin{equation}
\vspace{-1ex}
\nabla\nabla \approx -\,\kappa^2\,\mathbf{k}_{\mathrm{T},\ell}\mathbf{k}_{\mathrm{T},\ell}\trans.
\end{equation}

\noindent Leveraging these approximations, equation~\eqref{eq:Greens_function} on the \ac{TX} side can be rewritten as
\begin{equation}
\label{eq:Greens_function_TX_approximated}
\bm{G}_{\mathrm{T}}(\mathbf{u}_\ell, \mathbf{s}) \approx 
\frac{e^{-j \kappa d_{\mathrm{T},\ell}} \, e^{j \kappa \mathbf{k}_{\mathrm{T},\ell}\trans \mathbf{s}} \, e^{-j \kappa \tilde{v}_{\mathrm{T},\ell} t}}
{4 \pi d_{\mathrm{T},\ell}} 
\left(\mathbf{I}_3 - \mathbf{k}_{\mathrm{T},\ell} \mathbf{k}_{\mathrm{T},\ell}\trans \right).
\end{equation}

\vspace{-2ex}
\subsection{Wideband Effects}

After invoking the far-field approximation, we turn to the wideband effects that remain relevant. Let the signal occupy a bandwidth $B$ around a carrier frequency $f_c$. Denote the corresponding angular frequency and wavelength by $\omega_c=2\pi f_c$ and $\lambda_c=c/f_c$, and define the center wavenumber $\kappa_c=\omega_c/c=2\pi/\lambda_c$. For a generic frequency $f\in[f_c-B/2,\,f_c+B/2]$, the instantaneous wavenumber is $\kappa=2\pi f/c$. The dependence on $\kappa$ appears both in the spatial phase across the aperture and in the Doppler term induced by motion. We next introduce two standard small-parameter approximations that replace $\kappa$ by $\kappa_c$ in those terms when bandwidth-related products are small.

\vspace{0.5ex}
\noindent\textbf{Low Bandwidth-Aperture Product:} This condition states that spatial wideband effects across the array are negligible, so a frequency-independent phase well describes the array response. Let $\mathbf{k}_{\mathrm{T},\ell}$ be the unit departure direction of the $\ell$-th path and $\mathbf{s}$ be a sensor position on the transmit aperture. When the product $B D / c \ll 1$, where $D$ is the aperture extent, differential propagation delays across the aperture are much smaller than the inverse bandwidth, such that beam-squinting does not occur. Under this condition, the following approximation holds \cite{8354789, 10495774}   
\begin{equation}
\label{eq:wideband_effect_1}
\mathrm{exp}( j \kappa \mathbf{k}_{\mathrm{T},\ell}\trans \mathbf{s}) \approx \mathrm{exp}( j \kappa_c \mathbf{k}_{\mathrm{T},\ell}\trans \mathbf{s}) 
= \mathrm{exp}\Big( j \frac{2\pi}{\lambda_c} \mathbf{k}_{\mathrm{T},\ell} \mathbf{s}\Big).
\end{equation}

In other words, the spatial phase is evaluated at the center frequency for all tones across the band.

\vspace{0.5ex}
\noindent \textbf{Low Time-Bandwidth Product:} This condition yields a frequency-independent Doppler frequency. Since the Doppler frequency scales with $\kappa$, its variation across the band over an observation time $t$ is small when $(B\,\tilde{v}_{\mathrm{T},\ell}\,t)/c \ll 1$. Equivalently, over the processing bandwidth, the Doppler phase error caused by frequency spread is negligible. In this regime, we have the following approximation \cite{7968464, 9529026} \vspace{-0.5ex}
\begin{equation}
\label{eq:wideband_effect_2}
\mathrm{exp}( -j \kappa \mathbf{k}_{\mathrm{T},\ell}\trans\mathbf{v}_{\mathrm{T},\ell} t) \approx  \mathrm{exp}( -j \kappa_c \tilde{v}_{\mathrm{T},\ell} t) 
=  \exp( -j 2\pi \nu_{\mathrm{T},\ell} t),
\end{equation}
where $\nu_{\mathrm{T},\ell}\triangleq \tilde{v}_{\mathrm{T},\ell}/\lambda_c$ is the Doppler frequency at the carrier.

Substituting the low-bandwidth-aperture and low-time-bandwidth approximations into \eqref{eq:Greens_function_TX_approximated} gives
\vspace{-0.5ex}
\begin{align}
\vspace{-1ex}
\label{eq:Greens_function_TX_approximated_wideband}
\bm{G}_{\mathrm{T}}(\mathbf{u}_\ell, \mathbf{s})
&\approx \frac{ e^{-j 2\pi f_c \tau_{\mathrm{T},\ell}}
e^{j \tfrac{2\pi}{\lambda_c} \mathbf{k}_{\mathrm{T},\ell}\trans \mathbf{s}}
e^{-j 2\pi \nu_{\mathrm{T},\ell} t}}
{4 \pi d_{\mathrm{T},\ell}} (\mathbf{I}_3 - \mathbf{k}_{\mathrm{T},\ell}\mathbf{k}_{\mathrm{T},\ell}\trans).
\vspace{-1ex}
\end{align}

By the same reasoning, the receiver-side Green function is 
\vspace{-1ex}
\begin{align}
\vspace{-1ex}
\label{eq:Greens_function_RX_approximated_wideband}
\bm{G}_{\mathrm{R}}(\mathbf{r}, \mathbf{u}_\ell)
& \!\approx\! \frac{ e^{-j 2\pi f_c \tau_{\mathrm{R},\ell}}
e^{j \tfrac{2\pi}{\lambda_c} \mathbf{k}_{\mathrm{R},\ell}\trans \mathbf{r}}
e^{-j 2\pi \nu_{\mathrm{R},\ell} t}}
{4 \pi d_{\mathrm{R},\ell}}
\big(\mathbf{I}_3 - \mathbf{k}_{\mathrm{R},\ell}\mathbf{k}_{\mathrm{R},\ell}\trans\big),
\vspace{-1ex}
\end{align}
\vspace{-3ex}

\noindent where $\tau_{\mathrm{R},\ell}$ and $\nu_{\mathrm{R},\ell}$ are defined analogously. 

Combining these terms, the frequency-domain channel in \eqref{eq:Freq-domain-channel_response_w_scattering_matrix} becomes
\vspace{-1ex}
\begin{equation}
\label{eq:final_channel_model_H}
\vspace{-1ex}
\bm{H}(\mathbf{r},\mathbf{s})
= \sum_{\ell=1}^{L} h_\ell \bm{\Xi}_\ell
e^{-j 2\pi f_c \tau_\ell}
e^{-j 2\pi \nu_\ell t}
e^{j \tfrac{2\pi}{\lambda_c}\, \mathbf{k}_{\mathrm{R},\ell}\trans \mathbf{r}}
e^{j \tfrac{2\pi}{\lambda_c}\, \mathbf{k}_{\mathrm{T},\ell}\trans \mathbf{s}},
\end{equation}
where $\tau_\ell \triangleq \tau_{\mathrm{R},\ell}+\tau_{\mathrm{T},\ell}$ and $\nu_\ell \triangleq \nu_{\mathrm{R},\ell}+\nu_{\mathrm{T},\ell}$.

The polarization operator $\bm{\Xi}_\ell \in \mathbb{C}^{3\times 3}$ composes the transverse projectors with the per-path polarization transfer matrix
\begin{equation}
\vspace{-1ex}
\label{eq:polarization_matrix}
\bm{\Xi}_\ell \triangleq \big(\mathbf{I}_3 - \mathbf{k}_{\mathrm{R},\ell}\mathbf{k}_{\mathrm{R},\ell}\trans\big)\,
\mathbf{\Gamma}_\ell\,
\big(\mathbf{I}_3 - \mathbf{k}_{\mathrm{T},\ell}\mathbf{k}_{\mathrm{T},\ell}\trans\big),
\end{equation}
and the large-scale path gain is
\begin{equation}
\label{eq:channel_gain}
h_\ell \triangleq \frac{1}{\sqrt{L}}\,\frac{1}{(4\pi)^2 d_{\mathrm{R},\ell}\, d_{\mathrm{T},\ell}}.
\end{equation}

Finally, applying the inverse Fourier transform over frequency yields the delay-time representation
\vspace{-1ex}
\begin{align}
\vspace{-1ex}
\bm{H}(\mathbf{r},\mathbf{s}; \tau, t)
= \sum_{\ell=1}^{L} h_\ell \bm{\Xi}_\ell
\delta(\tau - \tau_\ell)
e^{-j 2\pi \nu_\ell t}
e^{j \tfrac{2\pi}{\lambda_c} \mathbf{k}_{\mathrm{R},\ell}\trans \mathbf{r}}
e^{j \tfrac{2\pi}{\lambda_c} \mathbf{k}_{\mathrm{T},\ell}\trans \mathbf{s}}.\nonumber \\[-3ex] \label{eq:final_channel_model_H_time-delay}
\end{align}

\vspace{-1ex}

Notice that equation \eqref{eq:final_channel_model_H_time-delay} now represents a novel \ac{DD} channel structure effectively extending \cite{MatzTWC2005,RouSPM2024} to incorporate the dependence on the arbitrary continuous surfaces $\mathbf{r}$ and $\mathbf{s}$, as opposed to conventional discrete antenna structures.

\vspace{-2ex}
\subsection{Application to Discrete Array Settings}

Notice that equation~\eqref{eq:final_channel_model_H_time-delay} defines a \ac{DD} channel model for antennas with arbitrary geometries and wave vectors.
To provide a tractable example and illustrate its generality, we consider a \ac{UPA} with antenna elements arranged in a rectangular grid on the $x\text{-}z$ plane. The transmit array consists of $N_\mathrm{T} = N_{\mathrm{T},x} \times N_{\mathrm{T},z}$ elements with spacing $d_{\mathrm{T},x}$ and $d_{\mathrm{T},z}$ along the $x$- and $z$-axes, respectively. Similarly, the receive array consists of $N_\mathrm{R} = N_{\mathrm{R},x} \times N_{\mathrm{R},z}$ elements with spacing $d_{\mathrm{R},x}$ and $d_{\mathrm{R},z}$.
Defining the element indices as $n_{\mathrm{T},x} \triangleq \{0, \ldots, N_{\mathrm{T},x}-1\}$, $n_{\mathrm{T},z} \triangleq \{0, \ldots, N_{\mathrm{T},z}-1\}$, $n_{\mathrm{R},x} \triangleq \{0, \ldots, N_{\mathrm{R},x}-1\}$, and $n_{\mathrm{R},z} \triangleq \{0, \ldots, N_{\mathrm{R},z}-1\}$, 
\begin{subequations}
\begin{equation}
\label{eq:UPA_s}
\mathbf{s} \triangleq [n_{\mathrm{T},x} d_{\mathrm{T},x},\, 0,\, n_{\mathrm{T},z} d_{\mathrm{T},z}]^\mathsf{T} \in \mathbb{C}^{3 \times 1},
\end{equation}
\begin{equation}
\label{eq:UPA_r}
\mathbf{r} \triangleq [n_{\mathrm{R},x} d_{\mathrm{R},x},\, 0,\, n_{\mathrm{R},z} d_{\mathrm{R},z}]^\mathsf{T} \in \mathbb{C}^{3 \times 1},
\end{equation}
\end{subequations}
and
\begin{subequations}
\label{eq:K_t_r_UPA}
\begin{equation}
\label{eq:K_t_UPA}
\mathbf{k}_{\mathrm{T},\ell} \triangleq 
[\sin\theta_{\mathrm{T},\ell} \cos\phi_{\mathrm{T},\ell},\, 
\sin\theta_{\mathrm{T},\ell} \sin\phi_{\mathrm{T},\ell},\, 
\cos\theta_{\mathrm{T},\ell}]^\mathsf{T} \in \mathbb{C}^{3 \times 1},
\end{equation}
\begin{equation}
\label{eq:K_r_UPA}
\mathbf{k}_{\mathrm{R},\ell} \triangleq 
[\sin\theta_{\mathrm{R},\ell} \cos\phi_{\mathrm{R},\ell},\, 
\sin\theta_{\mathrm{R},\ell} \sin\phi_{\mathrm{R},\ell},\, 
\cos\theta_{\mathrm{R},\ell}]^\mathsf{T} \in \mathbb{C}^{3 \times 1},
\end{equation}
\end{subequations}
where $\theta_{\mathrm{T},\ell}, \theta_{\mathrm{R},\ell} \in [0, \pi]$ are the elevation angles, and $\phi_{\mathrm{T},\ell}, \phi_{\mathrm{R},\ell} \in [0, 2\pi)$ are the azimuth angles of the $\ell$-th path.

Then, the phase terms can be calculated as
$$e^{j \frac{2\pi}{\lambda_c} \mathbf{k}_{\mathrm{T},\ell}^\mathsf{T} \mathbf{s}} = e^{j \frac{2\pi}{\lambda_c} \left( n_{\mathrm{T},x} d_{\mathrm{T},x} \sin\theta_{\mathrm{T},\ell} \cos\phi_{\mathrm{T},\ell} + n_{\mathrm{T},z} d_{\mathrm{T},z} \cos\theta_{\mathrm{T},\ell} \right)},$$
$$e^{j \frac{2\pi}{\lambda_c} \mathbf{k}_{\mathrm{R},\ell}^\mathsf{T} \mathbf{r}} = e^{j \frac{2\pi}{\lambda_c} \left( n_{\mathrm{R},x} d_{\mathrm{R},x} \sin\theta_{\mathrm{R},\ell} \cos\phi_{\mathrm{R},\ell} + n_{\mathrm{R},z} d_{\mathrm{R},z} \cos\theta_{\mathrm{R},\ell} \right)},$$
with the corresponding array response vectors given by
\begin{subequations}
\begin{equation}
\mathbf{a}_\mathrm{T}(\theta_{\mathrm{T},\ell}, \phi_{\mathrm{T},\ell}) = \mathbf{a}_{\mathrm{T},z}(\theta_{\mathrm{T},\ell}) \otimes \mathbf{a}_{\mathrm{T},x}(\theta_{\mathrm{T},\ell}, \phi_{\mathrm{T},\ell}) \in \mathbb{C}^{N_\mathrm{T} \times 1},
\end{equation}
\begin{equation}
\mathbf{a}_\mathrm{R}(\theta_{\mathrm{R},\ell}, \phi_{\mathrm{R},\ell}) = \mathbf{a}_{\mathrm{R},z}(\theta_{\mathrm{R},\ell}) \otimes \mathbf{a}_{\mathrm{R},x}(\theta_{\mathrm{R},\ell}, \phi_{\mathrm{R},\ell})  \in \mathbb{C}^{N_\mathrm{R} \times 1},
\end{equation}
\end{subequations}
where
\begin{subequations}
\vspace{-1ex}
\begin{align}
\mathbf{a}_{\mathrm{T},x}(\theta_{\mathrm{T},\ell}, \phi_{\mathrm{T},\ell}) &= [ 1, e^{j \frac{2\pi}{\lambda_c} d_{\mathrm{T},x} \sin\theta_{\mathrm{T},\ell} \cos\phi_{\mathrm{T},\ell}}, \nonumber \\ 
&\hspace{-2ex}\cdots, e^{j \frac{2\pi}{\lambda_c} (N_{\mathrm{T},x}-1) d_{\mathrm{T},x} \sin\theta_{\mathrm{T},\ell} \cos\phi_{\mathrm{T},\ell}} ]^\mathsf{T},
\end{align}
\vspace{-6ex}
\begin{align}
\mathbf{a}_{\mathrm{T},z}(\theta_{\mathrm{T},\ell}) &= [ 1, e^{j \frac{2\pi}{\lambda_c} d_{\mathrm{T},z} \cos\theta_{\mathrm{T},\ell}}, \nonumber \\
&\cdots, e^{j \frac{2\pi}{\lambda_c} (N_{\mathrm{T},z}-1) d_{\mathrm{T},z} \cos\theta_{\mathrm{T},\ell}} ]^\mathsf{T},
\end{align}    
\end{subequations}
with analogous definitions for the \ac{RX} side.

Finally, the discrete \ac{MIMO} channel $\bm{H}(t,\tau) \in \mathbb{C}^{3N_\mathrm{R} \times 3N_\mathrm{T}}$ with all polarization effects included for a \ac{UPA} case with the definition $\acute{\mathbf{A}}_\ell \triangleq \mathbf{a}_\mathrm{R}(\theta_{\mathrm{R},\ell}, \phi_{\mathrm{R},\ell}) \mathbf{a}_\mathrm{T}\herm(\theta_{\mathrm{T},\ell}, \phi_{\mathrm{T},\ell})$ can be given as
\begin{equation}
\bm{H}(t,\tau) = \sum_{\ell=1}^L h_\ell \delta(\tau - \tau_\ell) e^{-j 2 \pi \nu_\ell t} \left( \acute{\mathbf{A}}_\ell \otimes \bm{\Xi}_\ell \right),
\label{eq:UPA_final_form_disc}
\end{equation}
where it can be seen that, by ignoring the \ac{3D} polarization matrices $\bm{\Xi}_\ell$, the model coincides with prior work done on discrete \ac{MIMO} systems in \ac{DD} channels \cite{RanasingheARXIV2025}.

% Example for $\mathbf{k}_{\mathrm{T},l}$:
% \begin{equation}
%     \mathbf{k}_{\mathrm{T},l} = [\cos(\theta)\sin(\phi), \sin(\theta)\sin(\phi), \cos(\phi)]\trans
% \end{equation}

\vspace{-1ex}
\section{Signal Models for Continuous MIMO}

% \vspace{-2ex}

\subsection{Arbitrarily Modulated Signals}

%%%% Careful here;; multicolumn equation!!!
\begin{figure*}[t!]
\setcounter{equation}{29}
\normalsize
\begin{equation}
\bm{r}[n] 
= \sum_{\zeta=0}^\infty \Bigg[ \Bigg(\! \bigg( \sum_{\ell=1}^L \overbrace{\int_{\mathcal{S}_{\mathrm{R}}}\! \int_{\mathrm{S}_{\mathrm{T}}} h_\ell \bm{J}_{\mathrm{R}}\herm(\mathbf{r}) \bm{\Xi}_\ell \bm{J}_{\mathrm{T}}(\mathbf{s}) e^{j \frac{2\pi}{\lambda_c} \mathbf{k}_{\mathrm{R},\ell}\trans \mathbf{r}} e^{j \frac{2\pi}{\lambda_c} \mathbf{k}_{\mathrm{T},\ell}\trans \mathbf{s}} d \mathbf{s} d \mathbf{r} }^{\triangleq \check{\mathbf{H}}_\ell \in \mathbb{C}^{M \times M}}  \bigg) e^{-j 2\pi f_\ell \frac{n}{N}}  \delta[ \zeta - \zeta_\ell ] \Bigg)  \bm{c}[n - \zeta] \Bigg] \!\!+\! \bm{w}[n].
\label{eq:sampled_TD}
\end{equation}
\setcounter{equation}{26}
\hrulefill
\vspace{-3ex}
\end{figure*}

Let $\bm{c}(t) = [c_1(t),\dots,c_m(t),\dots,c_{M}(t)]^\mathsf{T} \in \mathbb{C}^{M \times 1}$ denote the $M$ transmit data streams from an arbitrary modulation (such as \ac{OFDM}, \ac{OTFS} or \ac{AFDM}) in the \ac{TD} at a time instance $t$. 
Then, the CAPA beamformed transmit signal from equation \eqref{eq:Freq-domain-transmit} can be expressed as \vspace{-1ex}
\begin{equation}
\bm{x}(\mathbf{s},t) = \sum_{m=1}^{M} \bm{j}_{\mathrm{T},m}(\mathbf{s}) c_m(t) = \bm{J}_{\mathrm{T}}(\mathbf{s}) \mathbf{c}(t) \in \mathbb{C}^{3 \times 1},
\vspace{-1ex}
\end{equation}
where $\bm{j}_{\mathrm{T},m}(\mathbf{s}) \in \mathbb{C}^{3 \times 1}$ denotes the transmit beamformer vector for the $m$-th data stream, which are concatenated in $\bm{J}_{\mathrm{T}}(\mathbf{s}) = [\bm{j}_{\mathrm{T},1}(\mathbf{s}),\dots,\bm{j}_{\mathrm{T},m}(\mathbf{s}),\dots,\bm{j}_{\mathrm{T},M}(\mathbf{s})] \in \mathbb{C}^{3 \times M}$. 

Leveraging the above, the received signal $\bm{y}(\mathbf{r}, t) \in \mathbb{C}^{3 \times 1}$ at point $\mathbf{r}$ on the \ac{RX} is given by 
\begin{align}
\bm{y}(\mathbf{r}, t) = & \int_{-\infty}^{+\infty} \!\!\int_{\mathrm{S}_{\mathrm{T}}}  \bm{H}(\mathbf{r},\mathbf{s}; \tau, t) \bm{x}(\mathbf{s}, t - \tau)  d \mathbf{s} d \tau + \bm{w}(\mathbf{r}, t) \nonumber \\
&\hspace{-9ex}= \int_{-\infty}^{+\infty} \!\! \int_{\mathrm{S}_{\mathrm{T}}}  \bm{H}(\mathbf{r},\mathbf{s}; \tau, t) \bm{J}_{\mathrm{T}}(\mathbf{s}) \mathbf{c}(t-\tau)  d \mathbf{s} d \tau + \bm{w}(\mathbf{r}, t),
\label{eq:cont_rx_TD_signal}
\end{align} 
where $\bm{w}(\mathbf{r}, t) \in \mathbb{C}^{3 \times 1}$ denotes the \ac{AWGN} at point $\mathbf{r}$ on the receive surface with spatially and temporally uncorrelated elements, each with zero mean and variance $\sigma_w^2$. 

Let $\bm{j}_{\mathrm{R},m}(\mathbf{r}) \in \mathbb{C}^{3 \times 1}$ denote the combiner to receive the $m$-th data stream with $\bm{J}_{\mathrm{R}}(\mathbf{r}) = [\bm{j}_{\mathrm{R},1}(\mathbf{r}),\dots,\bm{j}_{\mathrm{R},m}(\mathbf{r}),\dots,\bm{j}_{\mathrm{R},M}(\mathbf{r})] \in \mathbb{C}^{3 \times M}$.
Applying the combiner to the signal in equation \eqref{eq:cont_rx_TD_signal} yields the beamformed \ac{TD} receiver signal $\bm{r}(t) \in \mathbb{C}^{M \times 1}$, given by
\hfil

$~$
\vspace{-3.5ex}
\begin{align}
\bm{r}(t) &\!=\!\! \int_{\mathcal{S}_{\mathrm{R}}} \bm{J}_{\mathrm{R}}\herm(\mathbf{r}) \bm{y}(\mathbf{r}, t) d \mathbf{r} \nonumber \\
&\!=\!\! \int_{-\infty}^{+\infty} \!\!\!\bigg(\! \int_{\mathcal{S}_{\mathrm{R}}}\! \int_{\mathrm{S}_{\mathrm{T}}} \!\!\!\! \bm{J}_{\mathrm{R}}\herm(\mathbf{r}) \bm{H}(\mathbf{r},\mathbf{s}; \tau, t) \bm{J}_{\mathrm{T}}(\mathbf{s}) d \mathbf{s} d \mathbf{r}\! \bigg) \mathbf{c}(t-\tau) d \tau \nonumber \\
& \hspace{6.5ex}\!+\!\! \int_{\mathcal{S}_{\mathrm{R}}} \bm{J}_{\mathrm{R}}\herm(\mathbf{r}) \bm{w}(\mathbf{r}, t) d \mathbf{r}.
\label{eq:cont_rx_TD_signal_beamformed}
\end{align}  

\vspace{-1ex}
\textbf{\textit{Note:}} \textit{In this paper, \textbf{continuous \ac{MIMO}} refers to \ac{MIMO} systems equipped with \textbf{continuous arrays} but operating with \textbf{discrete signals}, as opposed to models with \textbf{continuous arrays} operating with \textbf{continuous signals}, which are beyond the scope of this article and will be considered in future work.}

Next, let $\bm{r}[n] \in \mathbb{C}^{M \times 1}$ and $\bm{c}[n] \in \mathbb{C}^{M \times 1}$, with $n \in \{ 0,\dots,N-1 \}$, be the finite sequences obtained after respectively sampling $\bm{r}(t)$ and $\bm{c}(t)$ at a sufficiently high sampling rate $F_S \triangleq \frac{1}{T_S}$ in Hz \cite{RanasingheICNC2025Oversampling} within a total bandwidth $B$.
Then, the discrete-time equivalent of the received signal in equation \eqref{eq:cont_rx_TD_signal_beamformed} can be obtained as seen in equation \eqref{eq:sampled_TD}, where $\zeta$ indicates the normalized discrete delay index, while $f_\ell \triangleq \frac{N\nu_\ell}{F_s}$ and $\zeta_\ell \triangleq \frac{\tau_\ell}{T_s}$ are the normalized Doppler shift and the associated normalized discrete delay index of each $\ell$-th path scatterer.

Utilizing the circular convolution leveraging a \ac{CP}, equation \eqref{eq:sampled_TD} can be re-expressed as \vspace{-0.5ex}
\setcounter{equation}{30}
\begin{equation}
\bm{r} = {\sum_{\ell=1}^L \Big(\check{\mathbf{H}}_\ell \otimes \mathbf{G}_\ell \Big)}  \bm{c} + \bm{w},
\label{eq:vectorized_TD_IO_kron}
\in \mathbb{C}^{N \times 1},
\vspace{-0.5ex}
\end{equation}
where $\bm{r} \triangleq [\bm{r}_1, \cdots, \bm{r}_{M}]\trans \in \mathbb{C}^{N M \times 1}$, $\bm{c} \triangleq [\bm{c}_1, \cdots, \bm{c}_{M}]\trans \in \mathbb{C}^{N M \times 1}$ and $\bm{w} \triangleq [\bm{w}_1, \cdots, \bm{w}_{M}]\trans \in \mathbb{C}^{N M \times 1}$ are the receive, transmit and \ac{AWGN} vectors, respectively, with $\bm{r}_{M} \triangleq [r_{M}[0],\cdots,r_{M}[n],\cdots,r_{M}[N-1]] \in \mathbb{C}^{1 \times N}$, $\bm{c}_{M} \triangleq [c_{M}[0],\cdots,c_{M}[n],\cdots,c_{M}[N-1]] \in \mathbb{C}^{1 \times N}$ and $\bm{w}_{M} \triangleq [w_{M}[0],\cdots,w_{M}[n],\cdots,w_{M}[N-1]] \in \mathbb{C}^{1 \times N}$.

In turn, each $\mathbf{G}_\ell \triangleq \mathbf{\Phi}_{\ell} \mathbf{Z}^{f_\ell} \mathbf{\Pi}^{\zeta_\ell} \in \mathbb{C}^{N \times N}$ comprises the diagonal prefix phase matrix $\boldsymbol{\Phi}_\ell  \!\in\! \mathbb{C}^{N \times N}$ with the chirp-cyclic prefix phase function $\phi(n)$ as in \cite{RanasingheTWC2025}, given by
\begin{equation}
\!\!\!\!\!\mathbf{\Phi}_{\ell} \!\triangleq\! \mathrm{diag}\big[e^{-j2\pi \cdot \phi(\zeta_\ell)}, \ldots, e^{-j2\pi \cdot\phi(1)}, \overbrace{1, \ldots, 1}^{\zeta_\ell \; \text{ones}}\big] \!\in\! \mathbb{C}^{N \times N}\!,\!\!\!
\label{eq:CCP_phase_matrix}
\end{equation}
the diagonal roots-of-unity matrix $\mathbf{Z} \!\in\! \mathbb{C}^{N \times N}$ given by 
\begin{equation}
\mathbf{Z} \triangleq \mathrm{diag}\big[e^{-j2\pi\frac{(0)}{N}}, \,\ldots\,, e^{-j2\pi\frac{(N\!-\!1)}{N}}\big] \in \mathbb{C}^{N \times N},
\label{eq:Z_matrix}
\end{equation}
which is taken to the $f_\ell$-th power, and the circular left-shift matrix $\mathbf{\Pi} \in \mathbb{C}^{N \times N}$ with elements
\begin{equation}
\label{eq:PiMatrix}
\pi_{i,j} \triangleq \delta_{i,j+1} + \delta_{i,j-(N-1)}\,\;\; \delta _{ij} \triangleq
\begin{cases}
0 & \text{if }i\neq j,\\
1 & \text{if }i=j.
\end{cases}
\vspace{-0.25ex}
\end{equation}

\vspace{-2ex}
\subsection{OFDM Signaling}

Let $\mathcal{D}$ denote an arbitrary complex constellation set of cardinality $D$ and average energy $E_\mathrm{S}$, which is associated with a given digital modulation scheme (e.g., \ac{QAM}). In \ac{OFDM}, multiple information vectors $\mathbf{x}_m \in \mathcal{C}^{N\times 1}$ with $m = \{ 1,\dots,M \}$, containing a total of $NM$ symbols, are modulated into the following transmit signal as
\begin{equation}
\label{eq:OFDM_modulation}
\bm{c}^\text{OFDM}_m \triangleq \mathbf{F}_N\herm  \mathbf{x}_m \in \mathbb{C}^{N \times 1},
\end{equation}
where $\mathbf{F}_N$ denotes the $N$-point normalized \ac{DFT} matrix. 

After undergoing circular convolution with the \ac{DD} channel and using a formulation similar to equation \eqref{eq:vectorized_TD_IO_kron}, the corresponding $NM$-element discrete-time received \ac{OFDM} signal can be written as \vspace{-1ex}
\begin{equation}
\label{eq:TD_OFDM_input_output}
\bm{r}^\text{OFDM} \triangleq \bar{\mathbf{H}}  \bm{c}^\text{OFDM} + \bm{w},% \in \mathbb{C}^{Nd_s \times 1},
\vspace{-1ex}
\end{equation}
where the $NM$-element vectors are defined as
\begin{equation}
\label{eq:OFDM_stacked_s}
\bm{c}^\text{OFDM} \triangleq 
\begin{bmatrix}
\bm{c}^\text{OFDM}_1 \\[-1ex]
\vdots \\
\bm{c}^\text{OFDM}_{M}
\end{bmatrix},\,\,
\bm{r}^\text{OFDM} \triangleq 
\begin{bmatrix}
\bm{r}^\text{OFDM}_1 \\[-1ex]
\vdots \\
\bm{r}^\text{OFDM}_{M}
\end{bmatrix}.
\end{equation}
\vspace{-1ex}

At the \ac{RX} side, applying \ac{OFDM} demodulation yields
\begin{equation}
\label{eq:OFDM_demodulation}
\mathbf{y}^\text{OFDM}_m \triangleq \mathbf{F}_N  \bm{r}^\text{OFDM}_m \in \mathbb{C}^{N \times 1},
\end{equation}
yielding the corresponding $NM$-element discrete-time signal
\begin{equation}
\label{eq:OFDM_input_output}
\mathbf{y}^\text{OFDM} = \bar{\mathbf{H}}^\text{OFDM}  \mathbf{x} + \bar{\mathbf{w}}^\text{OFDM}, 
%\in \mathbb{C}^{Nd_s \times 1},
\end{equation}

\noindent where $\bar{\mathbf{w}}^\text{OFDM} \in \mathbb{C}^{NM \times 1}$ is an equivalent \ac{AWGN} with the same statistics of $\bm{w}$, and $\bar{\mathbf{H}}^\text{OFDM} \in \mathbb{C}^{NM \times NM}$ is the effective \ac{OFDM} channel defined similar to $\bar{\mathbf{H}}$ in equation \eqref{eq:vectorized_TD_IO_kron} as
\begin{equation}
\label{eq:OFDM_effective_channel}
\bar{\mathbf{H}}^\text{OFDM}
\triangleq \sum_{\ell=1}^L \check{\mathbf{H}}_\ell \otimes \overbrace{( \mathbf{F}_N \mathbf{G}_\ell  \mathbf{F}_N\herm)}^{\triangleq\mathbf{G}_\ell^\text{OFDM} \in \mathbb{C}^{N \times N}} = \sum_{\ell=1}^L \check{\mathbf{H}}_\ell \otimes \mathbf{G}_\ell^\text{OFDM}.
\end{equation}

Notice that for the \ac{OFDM} case, the \ac{CP} phase matrices $\mathbf{\Phi}_{\ell}$'s appearing in equation \eqref{eq:vectorized_TD_IO_kron} reduce to identity matrices \cite{RouSPM2024}, i.e., $\phi(n) = 0$ in equation \eqref{eq:CCP_phase_matrix}, since there is no phase offset.

% ------- CAREFUL MOVING THIS! EQUATION ON TOP OF PAGE!!! -----------
%
%
\begin{figure*}[t!]
\setcounter{equation}{48}
\normalsize
\begin{equation}
\label{eq:AFDM_diagonal_CP_matrix_def}
\bm{\varPhi}_\ell \triangleq \text{diag}\bigg( [ \underbrace{e^{-\jmath 2\pi {c_1} (N^2-2N\zeta_\ell)}, e^{-\jmath 2\pi {c_1} (N^2-2N(\zeta_\ell-1))}, \dots, e^{-\jmath 2\pi {c_1} (N^2-2N)}}_{\zeta_\ell \; \text{terms}}, \underbrace{1, 1, \dots, 1, 1}_{N - \zeta_\ell \; \text{ones}}] \bigg) \in \mathbb{C}^{N \times N}
\vspace{-1ex}
\end{equation}
\hrulefill
\vspace{-3ex}
\setcounter{equation}{40}
\end{figure*}
%
% ------------------------------------------------------------------
%
\subsection{OTFS Signaling}

When \ac{OTFS} is used, multiple matrices $\mathbf{X}_m\in \mathbb{C}^{\tilde{M}\times \tilde{M}'}$ with $m = \{ 1,\dots,M \}$, containing a total of $\tilde{M} \tilde{M}' M$ symbols taken from an arbitrary complex constellation $\mathcal{D}$, are modulated as\footnote{For simplicity, we assume that all pulse-shaping operations utilize rectangular waveforms such that the corresponding sample matrices can be reduced to identity matrices.}
\vspace{-0.5ex}
\begin{equation}
\label{eq:TD_transmit_matrix_vectorized}
\!\bm{c}^\text{OTFS}_m \!\triangleq\! \text{vec}\big(\bm{C}_m\big)\! =\! (\mathbf{F}_{\tilde{M}'}\herm \otimes \mathbf{I}_{\tilde{M}})  \text{vec}\big( \mathbf{X}_m
\big) \in \mathbb{C}^{\tilde{M}\tilde{M}'\times 1},
\end{equation}
where $\text{vec}(\cdot)$ denotes matrix vectorization via column stacking and $\bm{C}_m$ is a \ac{TD} symbols' matrix obtained from\footnote{Equivalently, $\bm{C}_m$ can be obtained as the Heisenberg transform of the \ac{ISFFT} of $\mathbf{X}_m$, $i.e.$, $\bm{C}_m = \mathbf{F}_{\tilde{M}}\herm \mathbf{X}_\text{FT}^m$ with $\mathbf{X}_\text{FT}^m \triangleq \mathbf{F}_{\tilde{M}} \mathbf{X}_m \mathbf{F}_{\tilde{M}'}\herm \in \mathbb{C}^{\tilde{M}\times \tilde{M}'}$.} the \ac{IDZT} of $\mathbf{X}_m$ as \cite{Hadani_WCNC_2017}
\begin{equation}
\label{eq:TD_transmit_matrix}
\bm{C}_m = \mathbf{X}_m \mathbf{F}_{\tilde{M}'}\herm  \in \mathbb{C}^{\tilde{M}\times \tilde{M}'}.
\end{equation}

We highlight that the notation in equation \eqref{eq:TD_transmit_matrix_vectorized} is in line with the strategy described in \cite{Raviteja_TWC_2018}, whereby the \ac{OTFS} signals are first vectorized and then appended with a \ac{CP} of length $N_\mathrm{CP}$ in order to eliminate inter-frame interference, in similarity with \ac{OFDM}. 
Taking advantage of this similarity, and to allow for direct comparisons between the two waveforms, we shall hereafter set $\tilde{M}\cdot \tilde{M}' = N$.

After transmission over the \ac{DD} channel $\bar{\mathbf{H}}$ as shown in equation \eqref{eq:vectorized_TD_IO_kron}, the $NM$-element discrete-time received \ac{OTFS} signal can be modeled similar to equation \eqref{eq:TD_OFDM_input_output} as $\bm{r}^\text{OTFS} \triangleq \bar{\mathbf{H}}  \bm{c}^\text{OTFS} + \bm{w}$, where the $NM$-element vectors $\bm{c}^\text{OTFS}$ and $\bm{r}^\text{OTFS}$ are defined for \ac{OTFS} similar to equation \eqref{eq:OFDM_stacked_s}.

However, unlike \ac{OFDM}, the detection of the information symbols $\mathbf{X}_m$'s from the $\bm{r}^\text{OTFS}_m$ elements $\forall$$m=1,\ldots,M$ of $\bm{r}^\text{OTFS}$ requires reversing the vectorization and the \ac{IDZT} operations employed in the construction of the $M$ elements of $\bm{c}^\text{OTFS}$, resulting in a distinct effective channel. 
In particular, let $\bm{R}_m \triangleq \text{vec}^{-1}(\bm{r}^\text{OTFS}_m) \in \mathbb{C}^{\tilde{M} \times \tilde{M}'}$, with $\text{vec}^{-1}(\cdot)$ indicating the de-vectorization operation according to which a vector of size $\tilde{M}\tilde{M}' \times 1$ is reshaped into a matrix of size $\tilde{M} \times \tilde{M}'$, and consider the following \ac{DZT}\footnote{Equivalently, $\mathbf{Y}_m$ can be obtained as the \ac{SFFT} of the Wigner transform of $\bm{R}_m$: $\mathbf{Y}_\text{FT}^m \triangleq \mathbf{F}_{\tilde{M}} \bm{R}_m$, yielding $\mathbf{Y}_m = \mathbf{F}_{\tilde{M}}\herm \mathbf{Y}_\text{FT}^m \mathbf{F}_{\tilde{M}'}\in \mathbb{C}^{\tilde{M} \times \tilde{M}'}$.}
\begin{equation}
\label{eq:DD_rec_sig_after_SFFT}
\mathbf{Y}_m  =  \bm{R}_m \mathbf{F}_{\tilde{M}'} \in \mathbb{C}^{\tilde{M} \times \tilde{M}'}.
\end{equation}
The demodulated \ac{OTFS} signal at the \ac{RX} then becomes
\begin{equation}
\label{eq:DD_demodulation}
\mathbf{y}^\text{OTFS}_m \triangleq \text{vec}(\mathbf{Y}_m) = (\mathbf{F}_{\tilde{M}'} \otimes \mathbf{I}_{\tilde{M}})  \bm{r}^\text{OTFS}_m \in \mathbb{C}^{N\times 1},
\end{equation}
which can be compactly written, similar to equation \eqref{eq:OFDM_input_output}, as the $NM$-element discrete-time received signal
\begin{equation}
\label{eq:DD_input_output_relation}
\mathbf{y}^\text{OTFS} = \bar{\mathbf{H}}^\text{OTFS}  \mathbf{x} + \bar{\mathbf{w}}^\text{OTFS},
%\in \mathbb{C}^{Nd_s \times 1},
\end{equation}
where $\bar{\mathbf{w}}^\text{OTFS} \in \mathbb{C}^{NM \times 1}$ is an equivalent \ac{AWGN} with the same statistics as $\bm{w}$, while $\bar{\mathbf{H}}^\text{OTFS} \in \mathbb{C}^{NM \times NM}$ represents the effective \ac{OTFS} channel and is given by
\begin{eqnarray}
\bar{\mathbf{H}}^\text{OTFS} \triangleq \sum_{\ell=1}^L \check{\mathbf{H}}_\ell \otimes \overbrace{( (\mathbf{F}_{\tilde{M}'} \otimes \mathbf{I}_{\tilde{M}}) \mathbf{G}_\ell  (\mathbf{F}_{\tilde{M}'}\herm \otimes \mathbf{I}_{\tilde{M}}))}^{\triangleq\mathbf{G}_\ell^\text{OTFS} \in \mathbb{C}^{N \times N}} \nonumber&& \\
&&\hspace{-43ex}= \sum_{\ell=1}^L \check{\mathbf{H}}_\ell \otimes {\mathbf{G}_\ell^\text{OTFS}}.
\label{eq:OTFS_effective_channel}
\end{eqnarray}

Notice that similarly to the \ac{OFDM} case, the \ac{CP} phase matrices $\mathbf{\Theta}_p$'s reduce to identity matrices \cite{RouSPM2024}.
Comparing the expressions in equation \eqref{eq:OFDM_effective_channel} and equation \eqref{eq:OTFS_effective_channel}, one can appreciate how \cite{RouSPM2024}'s channel modeling approach elucidates both the similarity in form as well as the distinction in effect between the \ac{OFDM} and \ac{OTFS} waveforms in \ac{DD} channels.

\vspace{-1ex}
\subsection{AFDM Signaling}

The signal for transmission per information vector $\mathbf{x}_m$ when \ac{AFDM} waveform is used for the considered \ac{DD} \ac{MIMO} channel is given by  the \ac{IDAFT} as
\begin{equation}
\label{eq:AFDM_moduation}
\bm{c}^\text{AFDM}_m \triangleq \mathbf{\Lambda}_1\herm  \mathbf{F}_{N}\herm  \mathbf{\Lambda}_2\herm  \mathbf{x}_m \in \mathbb{C}^{N \times 1},
\end{equation}
where the $N\times N$ matrices $\mathbf{\Lambda}_i$ with $i=1,2$ are defined as
\begin{equation}
\label{eq:lambda_def}
\mathbf{\Lambda}_i \triangleq \text{diag}\big(\big[1, e^{-\jmath2\pi c_i 2^2}, \ldots, e^{-\jmath2\pi c_i (N-1)^2}\big]\big),
\end{equation}
where the first central chirp frequency $c_1$ is an optimally designed parameter based on the maximum Doppler channel statistics \cite{Bemani_TWC_2023}, while the second central chirp frequency $c_2$ is a free parameter that can be exploited for various functionalities.

It was shown in \cite{RouSPM2024} that, after going through a \ac{DD} channel, an \ac{AFDM} modulated symbol vector $\bm{c}^\text{AFDM}_m$ with the inclusion of a \ac{CPP} can be modeled similar to equation \eqref{eq:vectorized_TD_IO_kron} by replacing the \ac{CP} matrix $\mathbf{\Phi}_{\ell}$ in equation \eqref{eq:CCP_phase_matrix} with the \ac{CPP} matrix $\bm{\varPhi}_\ell$ given by equation \eqref{eq:AFDM_diagonal_CP_matrix_def} (top of this page). 
This implies that function $\phi(n)$ in equation \eqref{eq:CCP_phase_matrix} needs to be set as $\phi(n) = c_1 (N^2 - 2Nn)$. 
To this end, the $NM$-element discrete-time received \ac{AFDM} signal can be modeled similar to equation \eqref{eq:TD_OFDM_input_output} as $\bm{r}^\text{AFDM} \triangleq \bar{\mathbf{H}}  \bm{c}^\text{AFDM} + \bm{w}$, where the $NM$-element vectors $\bm{c}^\text{AFDM}$ and $\bm{r}^\text{AFDM}$ are defined for \ac{AFDM} similar to equation \eqref{eq:OFDM_stacked_s}. 

The \ac{AFDM} demodulation of each of the $\bm{r}^\text{AFDM}_m$ with $m\in\{1,\ldots,M\}$ elements of $\bm{r}^\text{AFDM}$ is obtained as
\setcounter{equation}{49}
\begin{equation}
\mathbf{y}^\text{AFDM}_m = \mathbf{\Lambda}_2  \mathbf{F}_{N}  \mathbf{\Lambda}_1  \bm{r}^\text{AFDM}_m \in \mathbb{C}^{N\times 1},
\label{eq:AFDM_demodulation}
\end{equation}
yielding the following expression for the $NM$-element discrete-time received signal:
\begin{equation}
\mathbf{y}^\text{AFDM} = \bar{\mathbf{H}}^\text{AFDM}  \mathbf{x} + \bar{\mathbf{w}}^\mathrm{AFDM},%\in \mathbb{C}^{Nd_s \times 1},
\label{eq:DAF_input_output_relation}
\end{equation}
where $\bar{\mathbf{w}}^\mathrm{AFDM} \in \mathbb{C}^{NM \times 1}$ is an equivalent \ac{AWGN} holding the same statistics with $\bm{w}$, and $\bar{\mathbf{H}}^\text{AFDM} \in \mathbb{C}^{NM \times NM}$ indicates the effective \ac{AFDM} channel given by
\begin{eqnarray}
\label{eq:AFDM_effective_channel}
\bar{\mathbf{H}}^\text{AFDM} \triangleq \sum_{\ell=1}^L \check{\mathbf{H}}_\ell \otimes \overbrace{( \mathbf{\Lambda}_2  \mathbf{F}_{N}  \mathbf{\Lambda}_1 \mathbf{G}_\ell \mathbf{\Lambda}_1\herm  \mathbf{F}_{N}\herm  \mathbf{\Lambda}_2\herm)}^{\mathbf{G}_\ell^\text{AFDM} \in \mathbb{C}^{N \times N}}&& \\
&&\hspace{-39ex}=  \sum_{\ell=1}^L \check{\mathbf{H}}_\ell \otimes {\mathbf{G}_\ell^\text{AFDM}}.\nonumber 
\end{eqnarray}

Clearly, equation \eqref{eq:AFDM_effective_channel} has the same structure of equations \eqref{eq:OFDM_effective_channel} and \eqref{eq:OTFS_effective_channel}, with the same holding for the \ac{MIMO} input-output relationships described by equations \eqref{eq:OFDM_input_output}, \eqref{eq:DD_input_output_relation} and \eqref{eq:DAF_input_output_relation}. 
This implies that signal processing techniques such as channel estimation can be designed to under a unified framework, applying to \ac{OFDM}, \ac{OTFS}, \ac{AFDM}, and similar waveforms.

Finally, for the sake of clarity, we emphasize that a ``conventional'' discrete \ac{DD}-\ac{MIMO} model -- $i.e.$, a discrete \ac{MIMO} version used in conjunction with the example model in equation \eqref{eq:UPA_final_form_disc} -- can be trivially extracted from the above.

For instance, for the \ac{OFDM}, \ac{OTFS}, and \ac{AFDM} waveforms, equations \eqref{eq:OFDM_effective_channel}, \eqref{eq:OTFS_effective_channel}, and \eqref{eq:AFDM_effective_channel}, would yield
\vspace{-1ex}
\begin{equation}
\label{eq:H_DD_MIMO}
\bar{\mathbf{H}}^\text{MIMO}\triangleq \sum_{\ell=1}^L h_\ell \left( \acute{\mathbf{A}}_\ell \otimes \bm{\Xi}_\ell \right) \!\otimes\! {\mathbf{G}_\ell^\text{MIMO}},
\vspace{-1ex}
\end{equation}
where the previous subscripts \ac{OFDM}, \ac{OTFS}, and \ac{AFDM}, are respectively represented by the generic subscript \ac{MIMO}.

In addition, the general representation for \ac{OFDM}, \ac{OTFS} and \ac{AFDM} can now be expressed as
\vspace{-1ex}
\begin{equation}
\mathbf{y} = \sum_{\ell=1}^L \Big(\check{\mathbf{H}}_\ell \otimes \bar{\mathbf{G}}_\ell \Big)  \mathbf{x} + \bar{\mathbf{w}},
\label{eq:General_IO_model}
\vspace{-1ex}
\end{equation}
where we omit all the waveform specific super/subscripts.

% \section{Wavenumber-Domain Transform}
\vspace{-2ex}
\section{Transmit and Receive Beamforming Design}

In this section, we aim to use the continuous beamformers $\bm{J}_{\mathrm{T}}(\mathbf{s})$ and $\bm{J}_{\mathrm{R}}(\mathbf{r})$ to maximize the Frobenius norm of $\check{\mathbf{H}}_\ell, \forall \ell$ in order to increase the receive power of a system leveraging continuous arrays, to obtain equivalents for \ac{MRC} used in discrete systems.

\vspace{-2ex}
\subsection{Problem Formulation}

Let us start by defining
\vspace{-1ex}
\begin{equation}
\label{eq:H_reform_opt}
\mathbf{H}(\mathbf{r}, \mathbf{s}) \triangleq \sum_{\ell=1}^L h_\ell \bm{\Xi}_\ell  e^{j \frac{2\pi}{\lambda_c} \mathbf{k}_{\mathrm{R},\ell}\trans \mathbf{r}} e^{j \frac{2\pi}{\lambda_c} \mathbf{k}_{\mathrm{T},\ell}\trans \mathbf{s}}.
\vspace{-1ex}    
\end{equation}
Then, the complete objective function can be compactly expressed as
\vspace{-2ex}
\begin{align}
\label{eq:full_optimization_problem_channel_coeff}
&\mathcal{P}1: \underset{\bm{J}_{\mathrm{T}}(\mathbf{s}),\bm{J}_{\mathrm{R}}(\mathbf{r})}{\text{max}} \; \Big|\Big| \overbrace{\int_{\mathcal{S}_{\mathrm{R}}}\! \int_{\mathrm{S}_{\mathrm{T}}} \bm{J}_{\mathrm{R}}\herm(\mathbf{r}) \mathbf{H}(\mathbf{r}, \mathbf{s}) \bm{J}_{\mathrm{T}}(\mathbf{s}) d \mathbf{s} d \mathbf{r}}^{\mathbf{O}\big(\bm{J}_{\mathrm{T}}(\mathbf{s}), \bm{J}_{\mathrm{R}}(\mathbf{r})\big) \in \mathbb{C}^{M \times M}}  \Big|\Big|_F^2\nonumber\\
&\;\;\;\;\;\;\text{\text{s}.\text{t}.}\;\;\;\;\; \int_{\mathcal{S}_\mathrm{T}} \big|\big| \bm{J}_{\mathrm{T}}(\mathbf{s}) \big|\big|^2 d\mathbf{s} \leq P_\mathrm{T}, \nonumber \\
&\;\;\;\;\;\;\;\;\;\;\;\;\;\;\;\, \int_{\mathcal{S}_\mathrm{R}} \big|\big| \bm{J}_{\mathrm{R}}(\mathbf{r}) \big|\big|^2 d\mathbf{r} = 1,
\end{align}
where $P_\mathrm{T}$ is the power constraint at the transmitter, the second constraint a normalized scaling factor applied at the receiver, and $|| \cdot ||_F$ denotes the Frobenius norm.

The problem stated in equation \eqref{eq:full_optimization_problem_channel_coeff} is a non-convex \textit{functional programming} problem, where \textit{functional} refers to a specific type of function that takes a function as its input and produces a scalar, namely a function of function.
This type of problem is generally challenging to solve due to the following reasons. 
First, the optimization variables $\bm{J}_{\mathrm{T}}(\mathbf{s})$ and $\bm{J}_{\mathrm{R}}(\mathbf{r})$ are not finite-sized
vectors or matrices, but continuous functions that can be viewed as infinite-dimensional vectors. 
Second, both the objective function and its constraints are defined in terms of integrals.
This kind of problem is typically solved by using commercial \ac{EM} simulation software, which may result in extremely high computational complexity. 
The Fourier-based approach \cite{ZhangJSAC2023}, is a \ac{SotA} alternative to address this problem recently. 
However, as has been shown in \cite{WangTWC2025,wang2025beamformingdesigncontinuousaperture}, the complexity of this approach is still high, as it requires a large number of basis
functions to approximate the functions. 
Additionally, this approximation makes it difficult to guarantee the optimality of the obtained solution. 

To address the above challenges, we propose a two-stage \ac{CoV}-based\footnotemark approach which can directly optimize the continuous functions $\bm{J}_{\mathrm{T}}(\mathbf{s})$ and $\bm{J}_{\mathrm{R}}(\mathbf{r})$ without approximation, while maintaining low computational complexity.

\footnotetext{The \ac{CoV} is a powerful tool for addressing such functional optimization problems \cite{Gelfand2000}.}

\subsection{Subproblem I: TX BF Formulation and Solution}

In order to obtain tractable solutions to $\mathcal{P}1$ in equation \eqref{eq:full_optimization_problem_channel_coeff}, let us first consider the continuous optimization over the transmit side by fixing $\bm{J}_{\mathrm{R}}(\mathbf{r})$ as
\begin{align}
\label{eq:full_optimization_problem_channel_coeff_TX}
&\mathcal{P}2: \underset{\bm{J}_{\mathrm{T}}(\mathbf{s})}{\text{max}} \; \Big|\Big| \int_{\mathcal{S}_{\mathrm{R}}}\! \int_{\mathrm{S}_{\mathrm{T}}} \bm{J}_{\mathrm{R}}\herm(\mathbf{r}) \mathbf{H}(\mathbf{r}, \mathbf{s}) \bm{J}_{\mathrm{T}}(\mathbf{s}) d \mathbf{s} d \mathbf{r}  \Big|\Big|_F^2\nonumber\\
&\;\;\;\;\;\;\text{\text{s}.\text{t}.}\;\;\;\;\; \int_{\mathcal{S}_\mathrm{T}} \big|\big| \bm{J}_{\mathrm{T}}(\mathbf{s}) \big|\big|^2 d\mathbf{s} \leq P_\mathrm{T}.
\end{align}

%%%%%%% From previous work, this formulation was seen to be good for sensing as well, so might be something to come back to later:))) %%%%%%%%%%%%

% Then, the optimization problem can be formulated as 
% %
% \begin{align}
% \label{eq:full_optimization_problem_channel_coeff}
% &\underset{\bm{J}_{\mathrm{T}}(\mathbf{s}),\bm{J}_{\mathrm{R}}(\mathbf{r})}{\text{max}} \; \sum_{\ell=1}^L \Big|\Big| \overbrace{\int_{\mathcal{S}_{\mathrm{R}}}\! \int_{\mathrm{S}_{\mathrm{T}}} h_\ell \bm{J}_{\mathrm{R}}\herm(\mathbf{r}) \bm{\Xi}_\ell \bm{J}_{\mathrm{T}}(\mathbf{s}) e^{j \frac{2\pi}{\lambda_c} \mathbf{k}_{\mathrm{R},\ell}\trans \mathbf{r}} e^{j \frac{2\pi}{\lambda_c} \mathbf{k}_{\mathrm{T},\ell}\trans \mathbf{s}} d \mathbf{r} d \mathbf{s}}^{\check{\mathbf{H}}_\ell}  \Big|\Big|_F^2\nonumber\\
% &\;\;\;\;\;\;\text{\text{s}.\text{t}.}\;\;\;\;\; \int_{\mathcal{S}_\mathrm{T}} | \bm{J}_{\mathrm{T}}(\mathbf{s}) |^2 d\mathbf{s} \leq P_\mathrm{T}, \nonumber \\
% &\;\;\;\;\;\;\;\;\;\;\;\;\;\;\;\, \int_{\mathcal{S}_\mathrm{R}} | \bm{J}_{\mathrm{R}}(\mathbf{r}) |^2 d\mathbf{r} \leq 1,
% \end{align}
% %
% where $P_\mathrm{T}$ and $1$ are the power constraints at the transmit and receive continuous beamformer.

In order to obtain a solution to $\mathcal{P}2$ defined in equation \eqref{eq:full_optimization_problem_channel_coeff_TX} based on \ac{CoV}, let us first recast it into an unconstrained optimization problem using the lemmas defined below.

\begin{lemma}[\textit{\ac{TX} Equivalent Power Constraint}]
\label{lemma:power_const}
The optimal solution to $\mathcal{P}2$ in equation \eqref{eq:full_optimization_problem_channel_coeff_TX} satisfies the power constraint with equality; i.e.,
\begin{equation}
\label{eq:equiv_pow_constraint}
\int_{\mathcal{S}_\mathrm{T}} \big|\big| \bm{J}_{\mathrm{T}}(\mathbf{s}) \big|\big|^2 d\mathbf{s} = P_\mathrm{T}.
\end{equation}

\begin{proof}
Let $\tilde{\bm{J}}_{\mathrm{T}}(\mathbf{s})$ denote a feasible solution to problem \eqref{eq:full_optimization_problem_channel_coeff_TX} that satisfies
\begin{equation}
\label{eq:proof_power_equi}
\tilde{P}_\mathrm{T} \triangleq \int_{\mathcal{S}_\mathrm{T}} \big|\big| \tilde{\bm{J}}_{\mathrm{T}}(\mathbf{s}) \big|\big|^2 d\mathbf{s} < P_\mathrm{T}.
\end{equation}

Next, by defining a scaling factor $\rho_t \triangleq P_\mathrm{T}/\tilde{P}_\mathrm{T}$ and a scaled solution $\bm{J}_{\mathrm{T}}(\mathbf{s}) \triangleq \sqrt{\rho_t} \tilde{\bm{J}}_{\mathrm{T}}(\mathbf{s})$, it can be readily shown that the maximum objective in \eqref{eq:full_optimization_problem_channel_coeff_TX} achieved by the scaled solution $\bm{J}_{\mathrm{T}}(\mathbf{s})$ must be higher than that achieved by the solution $\tilde{\bm{J}}_{\mathrm{T}}(\mathbf{s})$ since $\rho_t > 1$.

Additionally, it can also be shown that
\begin{equation}
\label{eq:proof_power_equi_shown}
\int_{\mathcal{S}_\mathrm{T}}\! \big|\big| \bm{J}_{\mathrm{T}}(\mathbf{s}) \big|\big|^2 d\mathbf{s}\! =\! \rho_t \int_{\mathcal{S}_\mathrm{T}} \big|\big| \tilde{\bm{J}}_{\mathrm{T}}(\mathbf{s}) \big|\big|^2 d\mathbf{s} \!=\! \rho_t \tilde{P}_\mathrm{T} \!=\! P_\mathrm{T}.
\end{equation}

The results in equation \eqref{eq:proof_power_equi_shown} implies that for any feasible solution to equation \eqref{eq:full_optimization_problem_channel_coeff_TX}, there exists a solution that achieves a larger maximum objective with a corresponding power equality constraint. 
The proof is therefore complete.
\end{proof}

\end{lemma}

\begin{lemma}[\textit{TX Unconstrained Equivalence Problem}]
\label{lemma:tx_opt}
Let $\bar{\bm{J}}_{\mathrm{T}}(\mathbf{s})$ denote an optimal solution to the functional maximization problem
\begin{equation}
\label{eq:full_optimization_problem_channel_coeff_TX_unconstrained}
\mathcal{P}3: \underset{\bm{J}_{\mathrm{T}}(\mathbf{s})}{\text{max}} \; \Big|\Big| \overbrace{\int_{\mathcal{S}_{\mathrm{R}}}\! \int_{\mathrm{S}_{\mathrm{T}}} \bm{J}_{\mathrm{R}}\herm(\mathbf{r}) \mathbf{H}(\mathbf{r}, \mathbf{s}) \bm{J}_{\mathrm{T}}(\mathbf{s}) d \mathbf{s} d \mathbf{r}}^{\triangleq \mathbf{A}_t[\bm{J}_{\mathrm{T}}] \in \mathbb{C}^{M \times M}} \Big|\Big|_F^2.
\end{equation}

An optimal solution to problem $\mathcal{P}2$ in equation \eqref{eq:full_optimization_problem_channel_coeff_TX} can then be expressed as
\begin{equation}
\label{eq:unconstrained_sol}
\bm{J}_{\mathrm{T}}(\mathbf{s}) = \sqrt{\frac{P_\mathrm{T}}{\int_{\mathcal{S}_\mathrm{T}} \big|\big| \bar{\bm{J}}_{\mathrm{T}}(\mathbf{s}) \big|\big|^2 d\mathbf{s}}} \bar{\bm{J}}_{\mathrm{T}}(\mathbf{s}).
\end{equation}

\begin{proof}
It is easy to show that the solution presented in equation \eqref{eq:unconstrained_sol} satisfies the equality power constraint in equation \eqref{eq:equiv_pow_constraint}, and ensures that the objective function in equation \eqref{eq:full_optimization_problem_channel_coeff_TX} always attains the same value as the objective function defined in equation \eqref{eq:full_optimization_problem_channel_coeff_TX_unconstrained}.
Therefore, leveraging the results obtained in Lemma \ref{lemma:power_const} and the fact that $\bar{\bm{J}}_{\mathrm{T}}(\mathbf{s})$ maximizes the objective function of problem $\mathcal{P}3$ in equation \eqref{eq:full_optimization_problem_channel_coeff_TX_unconstrained}, the solution $\bm{J}_{\mathrm{T}}(\mathbf{s})$ to problem $\mathcal{P}3$ in equation \eqref{eq:full_optimization_problem_channel_coeff_TX_unconstrained} must maximize the objective function $\mathcal{P}2$ in equation \eqref{eq:full_optimization_problem_channel_coeff_TX}.
This completes the proof.
\end{proof}

\end{lemma}

For ease of derivation, let us re-express the unconstrained optimization problem $\mathcal{P}3$ from equation \eqref{eq:full_optimization_problem_channel_coeff_TX_unconstrained} as
\begin{equation}
\label{eq:full_optimization_problem_channel_coeff_TX_unconstrained_cov}
\underset{\bm{J}_{\mathrm{T}}(\mathbf{s})}{\text{max}} \; E_t\big[ \bm{J}_{\mathrm{T}}(\mathbf{s}) \big] = \big|\big|\mathbf{A}_t[\bm{J}_{\mathrm{T}}]\big|\big|_F^2.
\end{equation}

This type of problem is termed a ``functional" optimization problem, which we can solve using the \ac{CoV} technique.
Some relevant definitions and theorems follow for ease of clarity.

% \begin{definition}[\textit{Functional Space}]
% \label{def:functional_def}
% Let $\mathcal{H} = L^2(\mathcal{S}_{\mathrm{T}}, \mathbb{C}^{3 \times M})$ be the Hilbert space of square-integrable functions from $\mathcal{S}_{\mathrm{T}}$ to $\mathbb{C}^{3 \times M}$ with inner product
% %
% \begin{equation}
% \inner{\bm{f}}{\bm{g}}_{\mathcal{H}} = \int_{\mathcal{S}_{\mathrm{T}}} \text{Tr}(\bm{f}\herm(\mathbf{s}) \bm{g}(\mathbf{s})) \, d\mathbf{s}.
% \end{equation}
% \end{definition}

\begin{definition}[\textit{Functional Space}]
\label{def:functional_def}
Let $\mathcal{H} = L^2(\mathcal{S}, \mathbb{C}^{3 \times M})$ be the Hilbert space of square-integrable functions from $\mathcal{S}$ to $\mathbb{C}^{3 \times M}$ with inner product
\vspace{-1ex}
\begin{equation}
\inner{\bm{f}}{\bm{g}}_{\mathcal{H}} = \int_{\mathcal{S}} \text{Tr}(\bm{f}\herm(\mathbf{a}) \bm{g}(\mathbf{a})) \, d\mathbf{s}.
\end{equation}
\end{definition}

% \begin{definition}[\textit{Functional Derivative}]
% \label{def:functional_derivative}
% For a functional $E: \mathcal{H} \to \mathbb{R}$, the functional derivative at $\bm{J}_{\mathrm{T}}(\mathbf{s})$ in direction $\bm{\eta}(\mathbf{s}) \in \mathbb{C}^{3 \times M}$ -- which is any arbitrary smooth function with each entry defined on $\mathcal{S}_{\mathrm{T}}$ -- is given by
% %
% \begin{equation}
% \delta E_t\big[\bm{J}_{\mathrm{T}}(\mathbf{s}); \bm{\eta}(\mathbf{s})\big] = \lim_{\epsilon \to 0} \frac{E_t\big[\bm{J}_{\mathrm{T}}(\mathbf{s}) + \epsilon \bm{\eta}(\mathbf{s})\big] - E_t\big[\bm{J}_{\mathrm{T}}(\mathbf{s})\big]}{\epsilon}.
% \end{equation}
% \end{definition}

\begin{definition}[\textit{Functional Derivative}]
\label{def:functional_derivative}
For a functional $E: \mathcal{H} \to \mathbb{R}$, the functional derivative at $\bm{J}(\mathbf{a})$ in direction $\bm{\eta}(\mathbf{a}) \in \mathbb{C}^{3 \times M}$, which is any arbitrary smooth function with each entry defined on $\mathcal{S}$, is given by
\begin{equation}
\!\delta E_t\big[\bm{J}(\mathbf{a}); \bm{\eta}(\mathbf{a})\big]\! =\! \lim_{\epsilon \to 0} \frac{E_t\big[\bm{J}(\mathbf{a}) + \epsilon \bm{\eta}(\mathbf{a})\big]\! - \!E_t\big[\bm{J}(\mathbf{a})\big]}{\epsilon}.\!
\end{equation}
\end{definition}

\begin{theorem}[\textit{TX Optimality Condition}]
\label{th:optimal_cond}
A necessary condition for $\bar{\bm{J}}_{\mathrm{T}}(\mathbf{s})$ to be optimal is
\begin{equation}
\delta E_t\big[\bar{\bm{J}}_{\mathrm{T}}(\mathbf{s}); \bm{\eta}(\mathbf{s})\big] = 0 ,\quad \forall \bm{\eta}(\mathbf{s}) \in \mathcal{H}.
\end{equation}

\begin{proof}
This is a standard result from calculus of variations.
If $\bar{\bm{J}}_{\mathrm{T}}(\mathbf{s})$ is a local extremum, then all the directional derivatives must vanish.
\end{proof}
\end{theorem}

Before computing the functional derivative of the function in equation \eqref{eq:full_optimization_problem_channel_coeff_TX_unconstrained_cov}, let us expand the squared norm leveraging the linearity of the integral of the integral operator as
\begin{align}
\label{eq:squared_norm_exp}
E_t\big[\bm{J}_{\mathrm{T}}(\mathbf{s}) + \epsilon \bm{\eta}(\mathbf{s})\big] &= \big|\big|\mathbf{A}_t[\bm{J}_{\mathrm{T}}] + \epsilon \mathbf{A}[\bm{\eta}] \big|\big|_F^2 \\
&\hspace{-17ex}= \big|\big|\mathbf{A}_t[\bm{J}_{\mathrm{T}}]\big|\big|_F^2 + 2\epsilon \Re\{\text{Tr}(\mathbf{A}_t[\bm{J}_{\mathrm{T}}]\herm \mathbf{A}[\bm{\eta}])\} + \epsilon^2 \big|\big|\mathbf{A}[\bm{\eta}]\big|\big|_F^2. \nonumber 
\end{align}

Then, using Definition \ref{def:functional_derivative}, the functional derivative can be expressed as
\vspace{-1ex}
\begin{align}
\label{eq:functional_derivative}
\delta E_t\big[\bm{J}_{\mathrm{T}}(\mathbf{s}); \bm{\eta}(\mathbf{s})\big] &= \frac{d}{d\epsilon} E_t\big[\bm{J}_{\mathrm{T}}(\mathbf{s}) + \epsilon \bm{\eta}(\mathbf{s})\big] \Big|_{\epsilon=0} \nonumber \\
&= 2 \Re\{\text{Tr}(\mathbf{A}_t[\bm{J}_{\mathrm{T}}]\herm \mathbf{A}[\bm{\eta}])\}.
\end{align}

Substituting the integral from equation \eqref{eq:full_optimization_problem_channel_coeff_TX_unconstrained} and leveraging the property that $\text{Tr}(\bm{A}\herm \bm{B}) = \text{Tr}(\bm{B}\herm \bm{A})$ gives us
\begin{align}
\delta E_t\big[\bm{J}_{\mathrm{T}}(\mathbf{s}); \bm{\eta}(\mathbf{s})\big] \\
&\hspace{-15.5ex}= 2 \Re\left\{\text{Tr}\left(\int_{\mathcal{S}_{\mathrm{R}}}\! \int_{\mathcal{S}_{\mathrm{T}}} \!\!\!\bm{\eta}\herm(\mathbf{s}') \bm{H}\herm(\mathbf{r}', \mathbf{s}') \bm{J}_{\mathrm{R}}(\mathbf{r}') \mathbf{A}_t[\bm{J}_{\mathrm{T}}] \, d\mathbf{s}' \, d\mathbf{r}' \right)\right\}\!\!.\nonumber 
\end{align}

Next, using Definition \ref{def:functional_def}, Theorem \ref{th:optimal_cond} and the fact that $\mathbf{A}_t[\bm{J}_{\mathrm{T}}]$ is independent of $\mathbf{s}'$, we get the condition
\begin{equation}
    \vspace{-1ex}
\label{eq:opt_cond_application}
\Re\left\{ \int_{\mathcal{S}_{\mathrm{T}}} \!\!\! \left\langle \bm{\eta}\herm(\mathbf{s}'), \int_{\mathcal{S}_{\mathrm{R}}}\!  \bm{H}\herm(\mathbf{r}', \mathbf{s}') \bm{J}_{\mathrm{R}}(\mathbf{r}') \mathbf{A}_t[\bm{J}_{\mathrm{T}}] \, d\mathbf{r}' \right\rangle_F d\mathbf{s}' \right\}\!\! = 0.
\end{equation}

\begin{lemma}[\textit{Fundamental Lemma of Calculus of Variations}]
\label{th:fundamental_lemma_cov}
If $\int_{\mathcal{S}} \inner{\bm{\eta}(\mathbf{a})}{\bm{f}(\mathbf{a})}_F d\mathbf{a} = 0$ for all $\bm{\eta} \in \mathcal{H}$, then $\bm{f}(\mathbf{a}) = \bm{0}$ almost everywhere \cite{Gelfand2000}.
\end{lemma}

Applying Lemma \ref{th:fundamental_lemma_cov} to equation \eqref{eq:opt_cond_application} yields
\begin{equation}
\label{eq:fundamental_cov_application}
\int_{\mathcal{S}_{\mathrm{R}}}\!  \bm{H}\herm(\mathbf{r}, \mathbf{s}) \bm{J}_{\mathrm{R}}(\mathbf{r}) \mathbf{A}_t[\bm{J}_{\mathrm{T}}] \, d\mathbf{r} = \bm{0}.
\end{equation}

Since we want to maximize $E_t\big[ \bm{J}_{\mathrm{T}}(\mathbf{s}) \big] = \|\mathbf{A}_t[\bm{J}_{\mathrm{T}}]\|_F^2$, we need $\mathbf{A}_t[\bm{J}_{\mathrm{T}}] \neq \bm{0}$. 
The condition then becomes an eigenvalue problem.
This implies that the optimal $\bar{\bm{J}}_{\mathrm{T}}(\mathbf{s})$ satisfies
\begin{align}
&\int_{\mathcal{S}_{\mathrm{R}}}\!\!\!\!\!  \bm{H}\herm(\mathbf{r}, \mathbf{s}) \bm{J}_{\mathrm{R}}(\mathbf{r}) d\mathbf{r} \! \int_{\mathcal{S}_{\mathrm{R}}}\! \int_{\mathrm{S}_{\mathrm{T}}}\!\!\!\! \bm{J}_{\mathrm{R}}\herm(\mathbf{r}') \mathbf{H}(\mathbf{r}', \mathbf{s}') \bm{J}_{\mathrm{T}}(\mathbf{s}') d \mathbf{s}' d \mathbf{r}' \nonumber \\
&\!=\! \lambda_t \bar{\bm{J}}_{\mathrm{T}}(\mathbf{s}).
\end{align}

Notice that the double integral $\tilde{\mathbf{A}} \triangleq \int_{\mathcal{S}_{\mathrm{R}}}\! \int_{\mathrm{S}_{\mathrm{T}}}\!\!\!\! \bm{J}_{\mathrm{R}}\herm(\mathbf{r}') \mathbf{H}(\mathbf{r}', \mathbf{s}') \bm{J}_{\mathrm{T}}(\mathbf{s}') d \mathbf{s}' d \mathbf{r}'$ is a constant matrix independent of $\mathbf{s}$.

\begin{theorem}[\textit{TX Matched Filter Solution}]
The optimal transmit beamformer for the Problem $\mathcal{P}2$ in equation \eqref{eq:full_optimization_problem_channel_coeff_TX} can be expressed as
\begin{equation}
\label{eq:MF_solution_TX}
\bm{J}_{\mathrm{T}}^\star(\mathbf{s}) = \bar{\mathbf{A}} \left( \int_{\mathcal{S}_{\mathrm{R}}}  \bm{H}\herm(\mathbf{r}, \mathbf{s}) \bm{J}_{\mathrm{R}}(\mathbf{r}) d\mathbf{r} \right) \frac{\tilde{\mathbf{A}}}{\lambda_t},
\end{equation}
where $\bar{\mathbf{A}}$ and $\tilde{\mathbf{A}}$ are constant matrices selected to satisfy the power constraint.

\begin{proof}
Since the double integral is a constant matrix $\tilde{\mathbf{A}}$, we have
\begin{equation}
\int_{\mathcal{S}_{\mathrm{R}}}\!\!\!\!\!  \bm{H}\herm(\mathbf{r}, \mathbf{s}) \bm{J}_{\mathrm{R}}(\mathbf{r}) d\mathbf{r} \tilde{\mathbf{A}} = \lambda_t \bar{\bm{J}}_{\mathrm{T}}(\mathbf{s}).
\end{equation}
This implies that
\begin{equation}
\bar{\bm{J}}_{\mathrm{T}}(\mathbf{s}) =  \left( \int_{\mathcal{S}_{\mathrm{R}}}\!\!\!\!\!  \bm{H}\herm(\mathbf{r}, \mathbf{s}) \bm{J}_{\mathrm{R}}(\mathbf{r}) d\mathbf{r} \right) \frac{\tilde{\mathbf{A}}}{\lambda_t}.
\end{equation}
Plugging in the power constraint from equation \eqref{eq:unconstrained_sol} results in 
\begin{equation}
\bm{J}_{\mathrm{T}}^\star(\mathbf{s}) =  \bar{\mathbf{A}} \left( \int_{\mathcal{S}_{\mathrm{R}}}  \bm{H}\herm(\mathbf{r}, \mathbf{s}) \bm{J}_{\mathrm{R}}(\mathbf{r}) d\mathbf{r} \right) \frac{\tilde{\mathbf{A}}}{\lambda_t},
\end{equation}
where
\begin{equation}
\bar{\mathbf{A}} \triangleq \sqrt{\frac{P_\mathrm{T}}{\int_{\mathcal{S}_\mathrm{T}} \big|\big| \left( \int_{\mathcal{S}_{\mathrm{R}}}\!\!\!  \bm{H}\herm(\mathbf{r}, \mathbf{s}) \bm{J}_{\mathrm{R}}(\mathbf{r}) d\mathbf{r} \right) \frac{\tilde{\mathbf{A}}}{\lambda_t} \big|\big|^2 d\mathbf{s}'}} .
\end{equation}

\end{proof}
\end{theorem}

\subsection{Subproblem II: RX BF Formulation and Solution}

Next, in a similar fashion to the TX BF optimization in the previous subsection, let us consider the continuous optimization over the receive side by fixing $\bm{J}_{\mathrm{T}}(\mathbf{s})$ to solve $\mathcal{P}1$ in equation \eqref{eq:full_optimization_problem_channel_coeff} as
\begin{align}
\label{eq:full_optimization_problem_channel_coeff_RX}
&\mathcal{P}4: \underset{\bm{J}_{\mathrm{R}}(\mathbf{r})}{\text{max}} \; \Big|\Big| \int_{\mathcal{S}_{\mathrm{R}}}\! \int_{\mathrm{S}_{\mathrm{T}}} \bm{J}_{\mathrm{R}}\herm(\mathbf{r}) \mathbf{H}(\mathbf{r}, \mathbf{s}) \bm{J}_{\mathrm{T}}(\mathbf{s}) d \mathbf{s} d \mathbf{r}  \Big|\Big|_F^2\nonumber\\
&\;\;\;\;\;\;\text{\text{s}.\text{t}.}\;\;\;\;\; \int_{\mathcal{S}_\mathrm{R}} \big|\big| \bm{J}_{\mathrm{R}}(\mathbf{r}) \big|\big|^2 d\mathbf{r} = 1.
\end{align}

In order to obtain a solution to $\mathcal{P}4$ defined in equation \eqref{eq:full_optimization_problem_channel_coeff_RX} based on \ac{CoV}, let us recast it into an unconstrained optimization problem using the lemmas defined below, similarly to what was done for the TX optimization.

\begin{remark}[\textit{\ac{RX} Power Constraint}]
\label{remark:power_const_RX}
Since the problem defined in equation \eqref{eq:full_optimization_problem_channel_coeff_RX} is by definition an equality power constraint due to the normalized scaling factor, there is no further equivalence to be demonstrated.
\end{remark}

\begin{lemma}[\textit{RX Unconstrained Equivalence Problem}]
\label{lemma:rx_opt}
Let $\bar{\bm{J}}_{\mathrm{R}}(\mathbf{r})$ denote an optimal solution to the functional maximization problem \vspace{-1ex}
\begin{equation}
\label{eq:full_optimization_problem_channel_coeff_RX_unconstrained}
\mathcal{P}5: \underset{\bm{J}_{\mathrm{R}}(\mathbf{r})}{\text{max}} \; \Big|\Big| \overbrace{\int_{\mathcal{S}_{\mathrm{R}}}\! \int_{\mathrm{S}_{\mathrm{T}}} \bm{J}_{\mathrm{R}}\herm(\mathbf{r}) \mathbf{H}(\mathbf{r}, \mathbf{s}) \bm{J}_{\mathrm{T}}(\mathbf{s}) d \mathbf{s} d \mathbf{r}}^{\triangleq \mathbf{A}_r[\bm{J}_{\mathrm{R}}] \in \mathbb{C}^{M \times M}} \Big|\Big|_F^2.
\end{equation}

An optimal solution to problem $\mathcal{P}4$ in equation \eqref{eq:full_optimization_problem_channel_coeff_RX} can then be expressed as
\begin{equation}
\label{eq:unconstrained_sol_RX}
\bm{J}_{\mathrm{R}}(\mathbf{r}) = \sqrt{\frac{1}{\int_{\mathcal{S}_\mathrm{R}} \big|\big| \bar{\bm{J}}_{\mathrm{R}}(\mathbf{r}) \big|\big|^2 d\mathbf{r}}} \bar{\bm{J}}_{\mathrm{R}}(\mathbf{r}).
\end{equation}

\begin{proof}
It is easy to show that the solution presented in equation \eqref{eq:unconstrained_sol_RX} satisfies the equality power constraint in equation \eqref{eq:full_optimization_problem_channel_coeff_RX}, and ensures that the objective function in equation \eqref{eq:full_optimization_problem_channel_coeff_RX} always attains the same value as the objective function defined in equation \eqref{eq:full_optimization_problem_channel_coeff_RX_unconstrained}.
Therefore, leveraging the results obtained in \textbf{Remark \ref{remark:power_const_RX}} and the fact that $\bar{\bm{J}}_{\mathrm{R}}(\mathbf{r})$ maximizes the objective function of problem $\mathcal{P}5$ in equation \eqref{eq:full_optimization_problem_channel_coeff_RX_unconstrained}, the solution $\bm{J}_{\mathrm{R}}(\mathbf{r})$ to problem $\mathcal{P}5$ in equation \eqref{eq:full_optimization_problem_channel_coeff_RX_unconstrained} must maximize the objective function $\mathcal{P}4$ in equation \eqref{eq:full_optimization_problem_channel_coeff_RX}.
This completes the proof.
\end{proof}

\end{lemma}

For ease of derivation, let us re-express the unconstrained optimization problem $\mathcal{P}5$ from equation \eqref{eq:full_optimization_problem_channel_coeff_RX_unconstrained} as
\begin{equation}
\label{eq:full_optimization_problem_channel_coeff_RX_unconstrained_cov}
\underset{\bm{J}_{\mathrm{R}}(\mathbf{r})}{\text{max}} \; E_r\big[ \bm{J}_{\mathrm{R}}(\mathbf{r}) \big] = \big|\big|\mathbf{A}_r[\bm{J}_{\mathrm{R}}]\big|\big|_F^2.
\end{equation}

\begin{theorem}[\textit{RX Optimality Condition}]
\label{th:optimal_cond_RX}
A necessary condition for $\bar{\bm{J}}_{\mathrm{R}}(\mathbf{r})$ to be optimal is 
\begin{equation}
\delta E_r\big[\bar{\bm{J}}_{\mathrm{R}}(\mathbf{r}); \bm{\eta}(\mathbf{r})\big] = 0 ,\quad \forall \bm{\eta}(\mathbf{r}) \in \mathcal{H}.
\end{equation}

\begin{proof}
This is a standard result from calculus of variations.
If $\bar{\bm{J}}_{\mathrm{R}}(\mathbf{r})$ is a local extremum, then all the directional derivatives must vanish.
\end{proof}
\end{theorem}

Before computing the functional derivative of the functional defined in equation \eqref{eq:full_optimization_problem_channel_coeff_RX_unconstrained_cov}, let us expand the squared norm leveraging the linearity of the integral operator as
\begin{align}
\label{eq:squared_norm_exp_RX}
E_r\big[\bm{J}_{\mathrm{R}}(\mathbf{r}) + \epsilon \bm{\eta}(\mathbf{r})\big] &= \big|\big|\mathbf{A}_r[\bm{J}_{\mathrm{R}}] + \epsilon \mathbf{A}[\bm{\eta}] \big|\big|_F^2  \\
&\hspace{-17ex}= \big|\big|\mathbf{A}_r[\bm{J}_{\mathrm{R}}]\big|\big|_F^2 + 2\epsilon \Re\{\text{Tr}(\mathbf{A}_r[\bm{J}_{\mathrm{R}}]\herm \mathbf{A}[\bm{\eta}])\} + \epsilon^2 \big|\big|\mathbf{A}[\bm{\eta}]\big|\big|_F^2. \nonumber
\end{align}

Then, using Definition \ref{def:functional_derivative}, the functional derivative for the RX side can be expressed as
\begin{align}
\label{eq:functional_derivative_RX}
\delta E_r\big[\bm{J}_{\mathrm{R}}(\mathbf{r}); \bm{\eta}(\mathbf{r})\big] &= \frac{d}{d\epsilon} E_r\big[\bm{J}_{\mathrm{R}}(\mathbf{r}) + \epsilon \bm{\eta}(\mathbf{r})\big] \Big|_{\epsilon=0} \nonumber \\
&= 2 \Re\{\text{Tr}(\mathbf{A}_r[\bm{J}_{\mathrm{R}}]\herm \mathbf{A}[\bm{\eta}])\}.
\end{align}

Substituting the integral from equation \eqref{eq:full_optimization_problem_channel_coeff_RX_unconstrained} and leveraging the property that $\text{Tr}(\bm{A}\herm \bm{B}) = \text{Tr}(\bm{B}\herm \bm{A})$ yields

\quad\\[-6ex]
\begin{align}
\label{eq:ffrrttgg}
\delta E_r\big[\bm{J}_{\mathrm{R}}(\mathbf{r}); \bm{\eta}(\mathbf{r})\big]  \\
&\hspace{-16ex}= 2 \Re\left\{\text{Tr}\left(\int_{\mathcal{S}_{\mathrm{R}}}\! \int_{\mathcal{S}_{\mathrm{T}}} \!\!\! \bm{J}_{\mathrm{T}}\herm(\mathbf{s}') \bm{H}\herm(\mathbf{r}', \mathbf{s}') \bm{\eta}(\mathbf{r}')  \mathbf{A}_r[\bm{J}_{\mathrm{R}}] \, d\mathbf{s}' \, d\mathbf{r}' \right)\right\}\!\!. \nonumber
\end{align}

Next, using the Frobenius inner product definition $\inner{\bm{A}}{\bm{B}}_F \triangleq \text{Tr}(\bm{A}\herm \bm{B})$, the expression in equation \eqref{eq:ffrrttgg} can be rewritten as
\begin{align}
\label{eq:ffrrttgg222}
\delta E_r\big[\bm{J}_{\mathrm{R}}(\mathbf{r}); \bm{\eta}(\mathbf{r})\big]  \\
&\hspace{-15.5ex}= 2 \Re\left\{\int_{\mathcal{S}_{\mathrm{R}}}\! \int_{\mathcal{S}_{\mathrm{T}}} \!\!\! \inner{\bm{\eta}(\mathbf{r}')}{\bm{H}(\mathbf{r}', \mathbf{s}') \bm{J}_{\mathrm{T}}(\mathbf{s}') \mathbf{A}_r\herm[\bm{J}_{\mathrm{R}}]}_F 
\, d\mathbf{s}' \, d\mathbf{r}' \right\}\!\!. \nonumber
\end{align}

Next, using Definition \ref{def:functional_def}, Theorem \ref{th:optimal_cond_RX} and the fact that $\mathbf{A}_r\herm[\bm{J}_{\mathrm{R}}]$ is independent of $\mathbf{r}'$, we get the condition
\begin{equation}
\label{eq:opt_cond_application_RX}
\Re\left\{ \int_{\mathcal{S}_{\mathrm{R}}} \!\!\! \left\langle \bm{\eta}(\mathbf{r}'), \int_{\mathcal{S}_{\mathrm{T}}}\!  \bm{H}(\mathbf{r}', \mathbf{s}') \bm{J}_{\mathrm{T}}(\mathbf{s}') \mathbf{A}_r\herm[\bm{J}_{\mathrm{R}}] \, d\mathbf{s}' \right\rangle_F d\mathbf{r}' \right\}\!\! = 0.
\end{equation}

Applying Lemma \ref{th:fundamental_lemma_cov} to equation \eqref{eq:opt_cond_application_RX} yields
\begin{equation}
\label{eq:fundamental_cov_application_RX}
\int_{\mathcal{S}_{\mathrm{T}}}\!  \bm{H}(\mathbf{r}, \mathbf{s}) \bm{J}_{\mathrm{T}}(\mathbf{s}) \mathbf{A}_r\herm[\bm{J}_{\mathrm{R}}] \, d\mathbf{s} = \bm{0}.
\end{equation}

Since we want to maximize $E_r\big[ \bm{J}_{\mathrm{R}}(\mathbf{r}) \big] = \|\mathbf{A}_r[\bm{J}_{\mathrm{R}}]\|_F^2$, we need $\mathbf{A}_r\herm[\bm{J}_{\mathrm{R}}] \neq \bm{0}$. 
The condition then becomes an eigenvalue problem.
This implies that the optimal $\bar{\bm{J}}_{\mathrm{R}}(\mathbf{r})$ satisfies
\begin{equation}
\label{eq:ImplicitB}
\int\limits_{\mathcal{S}_{\mathrm{T}}}\!\!\!  \bm{H}(\mathbf{r}, \mathbf{s}) \bm{J}_{\mathrm{T}}(\mathbf{s}) d\mathbf{s} \!\! \overbrace{\int\limits_{\mathcal{S}_{\mathrm{R}}}\! \int\limits_{\mathrm{S}_{\mathrm{T}}}\!\!\! \bm{J}_{\mathrm{T}}\herm(\mathbf{s}') \mathbf{H}\herm(\mathbf{r}'\!, \mathbf{s}') \bm{J}_{\mathrm{R}}(\mathbf{r}') d \mathbf{s}' d \mathbf{r}'}^{ \triangleq \tilde{\mathbf{B}}} \! \!=\!\! \lambda_r \bar{\bm{J}}_{\mathrm{R}}(\mathbf{r}),
\end{equation}
where we highlight that the matrix $\tilde{\mathbf{B}}$ is independent of $\mathbf{r}$.

\begin{theorem}[\textit{RX Matched Filter Solution}]
The optimal transmit beamformer for the Problem $\mathcal{P}4$ in equation \eqref{eq:full_optimization_problem_channel_coeff_RX} can be expressed as
\begin{equation}
\label{eq:MF_solution_RX}
\bm{J}_{\mathrm{R}}^\star(\mathbf{r}) = \bar{\mathbf{B}} \left( \int_{\mathcal{S}_{\mathrm{T}}}  \bm{H}(\mathbf{r}, \mathbf{s}) \bm{J}_{\mathrm{T}}(\mathbf{s}) d\mathbf{s} \right) \frac{\tilde{\mathbf{B}}}{\lambda_r},
\end{equation}
where $\bar{\mathbf{B}}$ and $\tilde{\mathbf{B}}$ are constant matrices selected to satisfy the power constraint.

\begin{proof}
Since the double integral in equation \eqref{eq:ImplicitB} is a constant matrix $\tilde{\mathbf{B}}$, we have
\begin{equation}
\int_{\mathcal{S}_{\mathrm{T}}}\!\!\!\!\!  \bm{H}(\mathbf{r}, \mathbf{s}) \bm{J}_{\mathrm{T}}(\mathbf{s}) d\mathbf{s} \tilde{\mathbf{B}} = \lambda_r \bar{\bm{J}}_{\mathrm{R}}(\mathbf{r}).
\end{equation}

This implies that
\begin{equation}
\bar{\bm{J}}_{\mathrm{R}}(\mathbf{r}) =  \left( \int_{\mathcal{S}_{\mathrm{T}}}\!\!\!\!\!  \bm{H}(\mathbf{r}, \mathbf{s}) \bm{J}_{\mathrm{T}}(\mathbf{s}) d\mathbf{s} \right)\frac{\tilde{\mathbf{B}}}{\lambda_r}.
\end{equation}

Plugging in the power constraint from equation \eqref{eq:unconstrained_sol_RX} yields 
\begin{equation}
\bm{J}_{\mathrm{R}}^\star(\mathbf{r}) =  \bar{\mathbf{B}} \left( \int_{\mathcal{S}_{\mathrm{T}}}  \bm{H}(\mathbf{r}, \mathbf{s}) \bm{J}_{\mathrm{T}}(\mathbf{s}) d\mathbf{s} \right) \frac{\tilde{\mathbf{B}}}{\lambda_r},
\end{equation}
where
\begin{equation}
\bar{\mathbf{B}} \triangleq \sqrt{\frac{1}{\int_{\mathcal{S}_\mathrm{R}} \big|\big| \left( \int_{\mathcal{S}_{\mathrm{T}}}\!\!  \bm{H}(\mathbf{r}, \mathbf{s}) \bm{J}_{\mathrm{T}}(\mathbf{s}) d\mathbf{s} \right)\frac{\tilde{\mathbf{B}}}{\lambda_r} \big|\big|^2 d\mathbf{r}'}}.
\end{equation}

\end{proof}
\end{theorem}

\vspace{-5ex}
\subsection{Implementation, Convergence and Complexity}

The proposed optimization procedure summarized in Algorithm \ref{alg:proposed_decoder} requires the computation of two integrals; namely, $\bm{J}_{\mathrm{T}}^\star(\mathbf{s})$ from equation \eqref{eq:MF_solution_TX} and $\bm{J}_{\mathrm{R}}^\star(\mathbf{r})$ from equation \eqref{eq:MF_solution_RX}, at each iteration.
These integrals can be computed using the \ac{GL} quadrature \cite{olver10}, in the form
\begin{equation}
\label{eq:Gauss_legendre_general}
\int_{\bar{a}}^{\bar{b}} \bar{\psi}(\bar{x}) \, d\bar{x} \approx \frac{\bar{b} - \bar{a}}{2} \sum_{\bar{m}=1}^{\bar{M}} \bar{\omega}_{\bar{m}} \bar{\psi} \left( \frac{\bar{b} - \bar{a}}{2} \bar{\theta}_{\bar{m}} + \frac{\bar{a} + \bar{b}}{2} \right)
\end{equation}
where $\bar{M}$ denotes the number of sample points, $\bar{\omega}_{\bar{m}}$ are the quadrature weights, and $\bar{\theta}_{\bar{m}}$ define the roots of the $\bar{M}$-th Legendre polynomial. 
Trivially, a larger value of $\bar{M}$ results in a higher approximation accuracy of the integral.

First, $\bm{J}_{\mathrm{T}}^\star(\mathbf{s})$ (without the constant matrices for brevity) can be computed as \vspace{-0.5ex}
\begin{align}
\label{eq:MF_solution_TX_Gauss_1}
\bm{J}_{\mathrm{T}}^\star(\mathbf{s}) 
&=  \int_{\mathcal{S}_{\mathrm{R}}} \bm{H}\herm(\mathbf{r}, \mathbf{s}) \bm{J}_{\mathrm{R}}(\mathbf{r}) \, d\mathbf{r} \nonumber \\
&=  \int_{-\frac{D_{\mathrm{R},z}}{2}}^{\frac{D_{\mathrm{R},z}}{2}} \int_{-\frac{D_{\mathrm{R},x}}{2}}^{\frac{D_{\mathrm{R},x}}{2}} \bm{H}\herm(r_x, r_z, \mathbf{s}) \bm{J}_{\mathrm{R}}(r_x, r_z) \, d r_x d r_z \nonumber \\
&\approx  \frac{D_{\mathrm{R},x} D_{\mathrm{R},z}}{4} \sum_{\bar{m}_x=1}^{\bar{M}_x} \sum_{\bar{m}_z=1}^{\bar{M}_z} \omega_{\bar{m}_x} \omega_{\bar{m}_z} \nonumber \\ 
&\hspace{5ex}\times \bm{H}\herm\left( \frac{D_{\mathrm{R},x} \bar{\theta}_{\bar{m}_x}}{2}, \frac{D_{\mathrm{R},z} \bar{\theta}_{\bar{m}_z}}{2}, \mathbf{s} \right) \nonumber \\ 
&\hspace{5ex}\times \bm{J}_{\mathrm{R}}\left( \frac{D_{\mathrm{R},x} \theta_{\bar{m}_x}}{2}, \frac{D_{\mathrm{R},z} \theta_{\bar{m}_z}}{2} \right).
\vspace{-0.5ex}
\end{align}

Note that $\bm{J}_{\mathrm{T}}^\star(\mathbf{s}) $ computed via equation \eqref{eq:MF_solution_TX_Gauss_1} using the \ac{GL} quadrature is a 4-D tensor since it has to be computed for each coordinate on the transmit surface; i.e., $\bm{J}_{\mathrm{T}}^\star(\mathbf{s}) \in \mathbb{C}^{3 \times M \times \bar{M}_x \times \bar{M}_z}$ with the subsequent per-element beamformer given by $\bm{J}_{\mathrm{T}}^\star(s_x, s_y) \in \mathbb{C}^{3 \times M}$.

Next, $\bm{J}_{\mathrm{R}}^\star(\mathbf{r})$ (without the constant matrices for brevity) can be computed as \vspace{-0.5ex}
\begin{align}
\label{eq:MF_solution_RX_Gauss_1}
\bm{J}_{\mathrm{R}}^\star(\mathbf{r}) 
&=  \int_{\mathcal{S}_{\mathrm{T}}} \bm{H}(\mathbf{r}, \mathbf{s}) \bm{J}_{\mathrm{T}}(\mathbf{s}) \, d\mathbf{s} \nonumber \\
&=  \int_{-\frac{D_{\mathrm{T},z}}{2}}^{\frac{D_{\mathrm{T},z}}{2}} \int_{-\frac{D_{\mathrm{T},x}}{2}}^{\frac{D_{\mathrm{T},x}}{2}} \bm{H}(\mathbf{r}, s_x, s_z) \bm{J}_{\mathrm{T}}(s_x, s_z) \, d s_x d s_z \nonumber \\
&\approx  \frac{D_{\mathrm{T},x} D_{\mathrm{T},z}}{4} \sum_{\bar{k}_x=1}^{\bar{K}_x} \sum_{\bar{k}_z=1}^{\bar{K}_z} \omega_{\bar{k}_x} \omega_{\bar{k}_z} \nonumber \\ 
&\hspace{5ex}\times \bm{H}\left( \mathbf{r}, \frac{D_{\mathrm{T},x} \bar{\theta}_{\bar{k}_x}}{2}, \frac{D_{\mathrm{T},z} \bar{\theta}_{\bar{k}_z}}{2} \right) \nonumber \\ 
&\hspace{5ex}\times \bm{J}_{\mathrm{T}}\left( \frac{D_{\mathrm{T},x} \theta_{\bar{k}_x}}{2}, \frac{D_{\mathrm{T},z} \theta_{\bar{k}_z}}{2} \right).
\vspace{-0.5ex}
\end{align}

Similarly, $\bm{J}_{\mathrm{R}}^\star(\mathbf{r}) $ computed via equation \eqref{eq:MF_solution_RX_Gauss_1} using the \ac{GL} quadrature is also a 4-D tensor since it has to be computed for each coordinate on the receive surface; i.e., $\bm{J}_{\mathrm{R}}^\star(\mathbf{r}) \in \mathbb{C}^{3 \times M \times \bar{K}_x \times \bar{K}_z}$ with the subsequent per-element beamformer given by $\bm{J}_{\mathrm{R}}^\star(r_x, r_y) \in \mathbb{C}^{3 \times M}$.

\vspace{-1.5ex}
\begin{algorithm}[H]
\caption{\ac{DDC MIMO} Optimization for Transmit and Receive Beamforming Design}
\label{alg:proposed_decoder}
\setlength{\baselineskip}{11pt}
\textbf{Input:} Iterations $i_{\mathrm{S}}$ and transmit power $P_\mathrm{T}$. \\
\textbf{Output:} $\bm{J}_{\mathrm{T}}^\star(\mathbf{s})$ and $\bm{J}_{\mathrm{R}}^\star(\mathbf{r})$. 
\vspace{-1.5ex} 
\begin{algorithmic}[1]  
\STATEx \hspace{-3.5ex}\hrulefill
\STATEx \hspace{-3.5ex}\textbf{Initialization}
\STATEx \hspace{-3.5ex} - Set $\bm{J}_{\mathrm{T}}^\star(\mathbf{s})$ and $\bm{J}_{\mathrm{R}}^\star(\mathbf{r})$ to all ones scaled by the respective \STATEx \hspace{-2ex} transmit and receive powers.
\vspace{-1ex}
\STATEx \hspace{-3.5ex}\hrulefill
\STATEx \hspace{-3.5ex}\textbf{DDC MIMO Optimization}
\STATEx \hspace{-3.5ex}\textbf{for} $i=1$ to $i_{\mathrm{S}}$ \textbf{do}:
\STATE Calculate the objective function $\mathbf{O}\big(\bm{J}_{\mathrm{T}}(\mathbf{s}), \bm{J}_{\mathrm{R}}(\mathbf{r})\big)$ from eq. \eqref{eq:full_optimization_problem_channel_coeff}, via the \ac{GL} quadrature in eq. \eqref{eq:GL_Q_objective_func}.
\STATE Assign $\tilde{\mathbf{A}} = \tilde{\mathbf{B}} = \mathbf{O}\big(\bm{J}_{\mathrm{T}}(\mathbf{s}), \bm{J}_{\mathrm{R}}(\mathbf{r})\big)$.
\STATE Set $\lambda_t = \lambda_r = \mathrm{max}\big(\mathrm{SVD}\big(\mathbf{O}\big(\bm{J}_{\mathrm{T}}(\mathbf{s}), \bm{J}_{\mathrm{R}}(\mathbf{r})\big)\big)\big)$.
\STATE Calculate $\bm{J}_{\mathrm{T}}^\star(\mathbf{s})$ from eq. \eqref{eq:MF_solution_TX}, via the \ac{GL} quadrature in eq. \eqref{eq:MF_solution_TX_Gauss_1}.
\STATE Calculate $\bm{J}_{\mathrm{R}}^\star(\mathbf{r})$ from eq. \eqref{eq:MF_solution_RX}, via the \ac{GL} quadrature in eq. \eqref{eq:MF_solution_RX_Gauss_1}.
\STATEx \hspace{-3.5ex}\textbf{end for}

\end{algorithmic}
\end{algorithm}

Finally, the same procedure can be applied to calculate $\mathbf{O}\big(\bm{J}_{\mathrm{T}}(\mathbf{s}), \bm{J}_{\mathrm{R}}(\mathbf{r})\big)$ in equation \eqref{eq:full_optimization_problem_channel_coeff} as 
\vspace{-1ex}
\begin{align}
\label{eq:GL_Q_objective_func}
&\int_{\mathcal{S}_{\mathrm{R}}}\! \int_{\mathrm{S}_{\mathrm{T}}} \bm{J}_{\mathrm{R}}\herm(\mathbf{r}) \mathbf{H}(\mathbf{r}, \mathbf{s}) \bm{J}_{\mathrm{T}}(\mathbf{s}) d \mathbf{s} d \mathbf{r} \nonumber \\
&= \int_{-\frac{D_{\mathrm{R},z}}{2}}^{\frac{D_{\mathrm{R},z}}{2}} \int_{-\frac{D_{\mathrm{R},x}}{2}}^{\frac{D_{\mathrm{R},x}}{2}} 
\int_{-\frac{D_{\mathrm{T},z}}{2}}^{\frac{D_{\mathrm{T},z}}{2}} \int_{-\frac{D_{\mathrm{T},x}}{2}}^{\frac{D_{\mathrm{T},x}}{2}}
\bm{J}_{\mathrm{R}}\herm(r_x, r_z) \nonumber \\
&\hspace{3ex}\times \bm{H}(r_x, r_z, s_x, s_z) \bm{J}_{\mathrm{T}}(s_x, s_z) \, d s_x d s_z d r_x d r_z \nonumber \\
&\approx \frac{D_{\mathrm{R},x} D_{\mathrm{R},z} D_{\mathrm{T},x} D_{\mathrm{T},z}}{16} \sum_{\bar{m}_x=1}^{\bar{M}_x} \sum_{\bar{m}_z=1}^{\bar{M}_z} \sum_{\bar{k}_x=1}^{\bar{K}_x} \sum_{\bar{k}_z=1}^{\bar{K}_z} \nonumber \\
&\hspace{3ex}\times \omega_{\bar{m}_x} \omega_{\bar{m}_z} \omega_{\bar{k}_x} \omega_{\bar{k}_z} \bm{J}_{\mathrm{R}}\herm\Big( \frac{D_{\mathrm{R},x} \theta_{\bar{m}_x}}{2}, \frac{D_{\mathrm{R},z} \theta_{\bar{m}_ z}}{2} \Big) \nonumber \\
&\hspace{3ex}\times \bm{H}\Big( \frac{D_{\mathrm{R},x} \bar{\theta}_{\bar{m}_x}}{2}, \frac{D_{\mathrm{R},z} \bar{\theta}_{\bar{m}_z}}{2}, \frac{D_{\mathrm{T},x} \bar{\theta}_{\bar{k}_x}}{2}, \frac{D_{\mathrm{T},z} \bar{\theta}_{\bar{k}_z}}{2} \Big) \nonumber \\
&\hspace{3ex}\times \bm{J}_{\mathrm{T}}\Big( \frac{D_{\mathrm{T},x} \theta_{\bar{k}_x}}{2}, \frac{D_{\mathrm{T},z} \theta_{\bar{k}_z}}{2} \Big).
\end{align}

\vspace{-1ex}
To analyze the convergence of the proposed Algorithm \ref{alg:proposed_decoder}, let us recall that the objective
function in \eqref{eq:full_optimization_problem_channel_coeff} can be expressed as $\mathbf{O}\big(\bm{J}_{\mathrm{T}}(\mathbf{s}), \bm{J}_{\mathrm{R}}(\mathbf{r})\big)$. 
For two consecutive iterations, we have
\vspace{-2ex}
\begin{equation}
% \vspace{-1ex}
\mathbf{O}\big(\bm{J}_{\mathrm{T}}^{(i+1)}(\mathbf{s}), \bm{J}_{\mathrm{R}}^{(i+1)}(\mathbf{r})\big) \overset{(a)}{\geq} \mathbf{O}\big(\bm{J}_{\mathrm{T}}^{(i)}(\mathbf{s}), \bm{J}_{\mathrm{R}}^{(i)}(\mathbf{r})\big),
\vspace{-1ex}
\end{equation}
where step $(a)$ follows from the global optimality described in Lemma \ref{lemma:tx_opt} and Lemma \ref{lemma:rx_opt}.
Since the objective function is bounded from above, the strict convergence of the proposed Algorithm \ref{alg:proposed_decoder} is guaranteed.
For a numerical validation, we also provide a convergence plot as portrayed in Fig. \ref{fig:convergence}.

The main computational complexity of the proposed Algorithm \ref{alg:proposed_decoder} arises as follows: Step 1, 4 and 5 incur $\mathcal{O}(\bar{M}_x \bar{M}_z \bar{K}_x \bar{K}_z)$ at each iteration and Step 3 incurs $\mathcal{O}(M^2)$ due to the SVD at each iteration.

The complete optimization procedure with the corresponding implementation steps is summarized in Algorithm \ref{alg:proposed_decoder}.

\vspace{-1ex}
\section{Performance Analysis}

Unless otherwise specified, the system parameters listed in Table \ref{tab:simulation_parameters} are persistently used throughout this section.
For the algorithmic parameters, the maximum iterations $i_{\mathrm{S}}$ were set to 20 and the \ac{GL} quadrature sample points were set to $\bar{M}_x = \bar{M}_z = \bar{K}_x = \bar{K}_z = 10$.
For performance comparisons, we mainly focus on two main \ac{SotA} methods described below.

\noindent \textbf{Classical DD MIMO:} A conventional discrete \ac{MIMO} setup with discrete \acp{UPA} defined as in equation \eqref{eq:UPA_final_form_disc}, and solved via the classical singular value decomposition.
In order to draw fair comparisons, the initial receive power of this setup is set to be identical to that of the proposed continuous case.

\noindent\textbf{Conventional MIMO:} Following \cite{WangTWC2025,SanguinettiTWC2023}, we can define conventional spatially discrete antenna arrays, where the continuous surfaces $\mathcal{S}_\mathrm{T}$ and $\mathcal{S}_\mathrm{R}$ are occupied with discrete antennas with effective aperture $A_d = \frac{\lambda^2_c}{4\pi}$ and antenna spacing $d_t = d_r = \frac{\lambda}{2}$.

\begin{table}[H]
\centering
\caption{System Parameters}
\label{tab:simulation_parameters}
\begin{tabular}{|c|c|c|}
\hline
\textbf{Parameter} & \textbf{Symbol} & \textbf{Value} \\
\hline
Carrier Frequency & $f_c$ & 2.4 GHz \\
\hline
Carrier Wavelength & $\lambda_c$ & 0.1249 m \\
\hline
System Bandwidth & $B$ & 1 MHz \\
\hline
Sampling Frequency & $F_S$ & 1 MHz \\
\hline
Number of Subcarriers & $N$ & 64 \\
\hline
Total TX and RX antennas & $N_\mathrm{T}$ & 81, 289, 1089 \\
\hline
Total RF chains & $M$ & 10 \\
\hline
Number of Channel Scatterers & $L$ & 5 \\
\hline
Maximum Range & $R_\text{max}$ & 1500 m \\
\hline
Maximum Velocity & $V_\text{max}$ & 122 m/s \\
\hline
Aperture Size & $A_\mathrm{T}, A_\mathrm{R}$ & 0.25 m$^2$ \\
\hline
\end{tabular}
\end{table}
\vspace{2ex}

The location of the \ac{TX} $(n_{\mathrm{T},x}, n_{\mathrm{T},z})$-th antenna is 
\begin{equation}
\bar{\mathbf{s}}_{n_{\mathrm{T},x},n_{\mathrm{T},z}} \!=\!
\Big[
(n_{\mathrm{T},x} \!-\! 1)d_t \!-\! \frac{D_{\mathrm{T},x}}{2}, 0, 
(n_{\mathrm{T},z} \!-\! 1)d_t \!-\! \frac{D_{\mathrm{T},z}}{2}
\Big]\trans\!\!\!\!, 
\end{equation}
while the \ac{RX} $(n_{\mathrm{R},x}, n_{\mathrm{R},z})$-th antenna is 
\begin{equation}
\bar{\mathbf{r}}_{n_{\mathrm{R},x},n_{\mathrm{R},z}} \!=\!
\Big[
(n_{\mathrm{R},x} \!-\! 1)d_r \!-\! \frac{D_{\mathrm{R},x}}{2}, 0, 
(n_{\mathrm{R},z} \!-\! 1)d_r \!-\! \frac{D_{\mathrm{R},z}}{2}
\Big]\trans\!\!\!\!. 
\end{equation}

Therefore, the total numbers of antennas are
\begin{equation}
    N_\mathrm{T} = \left\lceil \frac{D_{\mathrm{T},x}}{d_t} \right\rceil \times \left\lceil \frac{D_{\mathrm{T},z}}{d_t} \right\rceil \;\text{and}\; N_\mathrm{R} = \left\lceil \frac{D_{\mathrm{R},x}}{d_r} \right\rceil \times \left\lceil \frac{D_{\mathrm{R},z}}{d_r} \right\rceil.
\end{equation}

Let $\mathcal{S}_{n_{\mathrm{T},x},n_{\mathrm{T},z}}$ denote the surface of the \ac{TX} $(n_{\mathrm{T},x}, n_{\mathrm{T},z})$-th antenna and $\mathcal{S}_{n_{\mathrm{R},x},n_{\mathrm{R},z}}$ denote the surface of the \ac{RX} $(n_{\mathrm{R},x}, n_{\mathrm{R},z})$-th antenna,
where $|\mathcal{S}_{n_{\mathrm{T},x},n_{\mathrm{T},z}}| = |\mathcal{S}_{n_{\mathrm{R},x},n_{\mathrm{R},z}}| = A_d$.
Then, leveraging equation \eqref{eq:full_optimization_problem_channel_coeff}, the effective discrete equivalent of the channel from a discrete \ac{TX} antenna to a discrete \ac{RX} antenna can be expressed as
\begin{align}
\tilde{\mathbf{H}}(n_{\mathrm{R},x}, n_{\mathrm{R},z}, n_{\mathrm{T},x}, n_{\mathrm{T},z}) &\triangleq \tfrac{1}{A_d^2} \!\!\!\!\!\!\int\limits_{\mathcal{S}_{n_{\mathrm{R},x},n_{\mathrm{R},z}}}\! \int\limits_{\mathcal{S}_{n_{\mathrm{T},x},n_{\mathrm{T},z}}} \!\!\!\!\!\!\!\!\mathbf{H}(\mathbf{r}, \mathbf{s}) d \mathbf{s} d \mathbf{r} \nonumber \\
&\hspace{-12ex}\approx A_d^2 \mathbf{H}(\bar{\mathbf{r}}_{n_{\mathrm{R},x},n_{\mathrm{R},z}}, \bar{\mathbf{s}}_{n_{\mathrm{T},x},n_{\mathrm{T},z}}) \in \mathbb{C}^{3 \times 3},
\end{align}
for which the following equivalent optimization problem holds:
\begin{align}
\label{eq:full_optimization_problem_channel_coeff_discrete}
&\underset{\substack{\bm{J}_{\mathrm{T}}(n_{\mathrm{T},x}, n_{\mathrm{T},z}),\\ \bm{J}_{\mathrm{R}}(n_{\mathrm{R},x}, n_{\mathrm{R},z})}}{\text{max}} \Big|\Big| \!\!\sum_{n_{\mathrm{R},x}=1}^{N_{\mathrm{R},x}} \sum_{n_{\mathrm{R},z}=1}^{N_{\mathrm{R},z}} \sum_{n_{\mathrm{T},x}=1}^{N_{\mathrm{T},x}} \sum_{n_{\mathrm{T},z}=1}^{N_{\mathrm{T},z}} \!\!\bm{J}_{\mathrm{R}}\herm(n_{\mathrm{R},x}, n_{\mathrm{R},z}) \nonumber \\ &\hspace{12ex}\times \tilde{\mathbf{H}}(n_{\mathrm{R},x}, n_{\mathrm{R},z}, n_{\mathrm{T},x}, n_{\mathrm{T},z}) \bm{J}_{\mathrm{T}}(n_{\mathrm{T},x}, n_{\mathrm{T},z})   \Big|\Big|_F^2\nonumber\\
&\;\;\;\;\;\;\text{\text{s}.\text{t}.}\;\;\;\;\; \int\limits_{\mathcal{S}_{n_{\mathrm{T},x},n_{\mathrm{T},z}}} \!\!\!\!\!\big|\big| \bm{J}_{\mathrm{T}}(\mathbf{s}) \big|\big|^2 d\mathbf{s} \leq P_\mathrm{T}, \nonumber \\
&\;\;\;\;\;\;\;\;\;\;\;\;\;\;\;\, \int\limits_{\mathcal{S}_{n_{\mathrm{R},x},n_{\mathrm{R},z}}} \!\!\!\!\!\big|\big| \bm{J}_{\mathrm{R}}(\mathbf{r}) \big|\big|^2 d\mathbf{r} = 1,
\end{align}
where the beamformers can be calculated in a similar manner to the procedure described in Algorithm \ref{alg:proposed_decoder}.

\begin{figure}[H]
\centering
\includegraphics[width=\columnwidth]{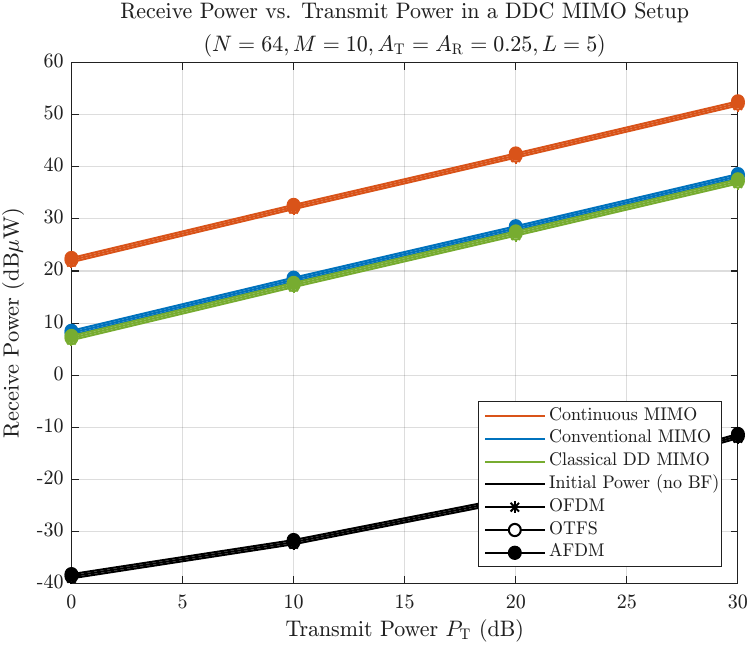}
\caption{Receive power vs. transmit power for the proposed DDC MIMO setup for \ac{OFDM}, \ac{OTFS} and \ac{AFDM}.}
\label{fig:tx_pow_vary}
\end{figure}

Our first set of results is presented in Fig.~\ref{fig:tx_pow_vary}, which compares the received power under varying transmit power $P_\mathrm{T}$ applied to the \ac{TX} beamformer for both conventional approaches and the proposed \ac{CAPA}-based method.
As shown in the figure, the proposed \ac{DDC MIMO} model yields a significant increase in received power compared to conventional \ac{MIMO} systems, owing to the exploitation of the entire aperture. 
Moreover, the proposed Algorithm~\ref{alg:proposed_decoder} achieves substantial gains over equal-power allocation across beamforming elements, as indicated by the black curves.
It is also noteworthy that the received power remains identical for \ac{OFDM}, \ac{OTFS}, and \ac{AFDM}, since the waveform-dependent matrices $\bar{\mathbf{G}}_\ell$ do not influence the power.

\begin{figure}[H]
\centering
\includegraphics[width=\columnwidth]{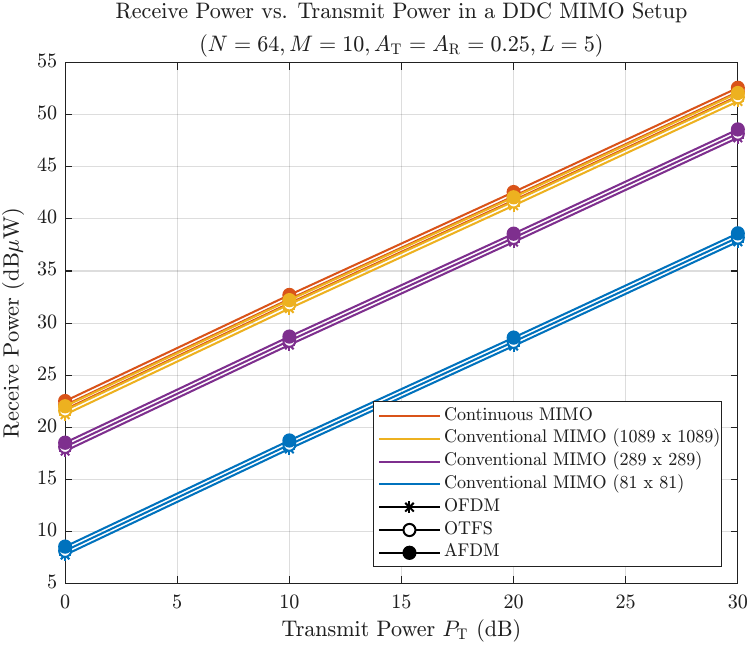}
\caption{Receive power vs. transmit power for the proposed DDC MIMO setup for \ac{OFDM}, \ac{OTFS} and \ac{AFDM} with an increasing number of discrete antennas.}
\label{fig:tx_pow_vary_anennas}
\end{figure}

\begin{figure}[H]
\includegraphics[width=\columnwidth]{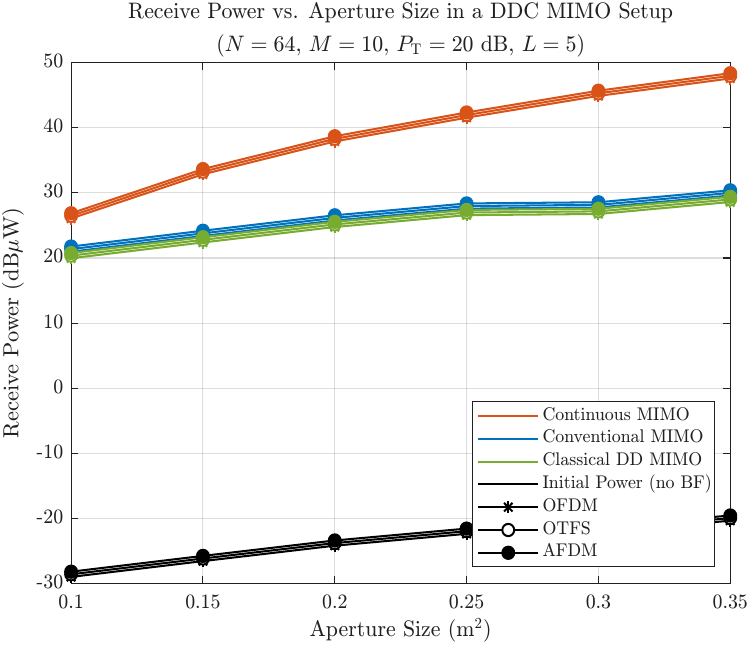}
\caption{Receive power vs. aperture size for the proposed DDC MIMO setup for \ac{OFDM}, \ac{OTFS} and \ac{AFDM}.}
\label{fig:aperture_vary}
\end{figure}

\begin{figure}[H]
\centering
\includegraphics[width=\columnwidth]{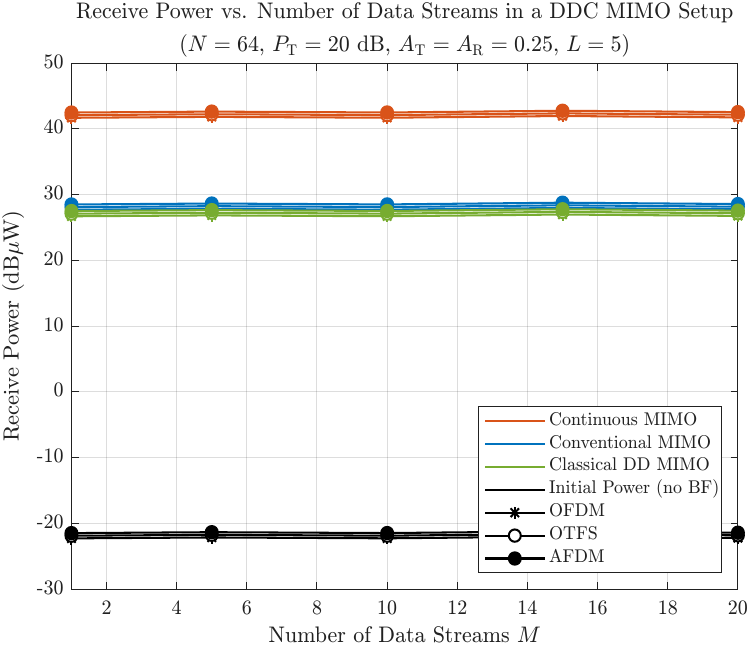}
\caption{Receive power vs. number of data streams for the proposed DDC MIMO setup for \ac{OFDM}, \ac{OTFS} and \ac{AFDM}.}
\label{fig:M_vary}
\end{figure}

Next, Fig.~\ref{fig:tx_pow_vary_anennas} illustrates the received power as a function of the number of \ac{TX} and \ac{RX} antennas in the conventional \ac{MIMO} setting.
As observed, the conventional cases approach the continuous case as the number of antennas increases, verifying that the proposed method accurately emulates a system with a very large number of antennas at a fraction of the complexity.

Fig.~\ref{fig:aperture_vary} presents results when the \ac{TX} and \ac{RX} aperture areas are increased with $A_\mathrm{T} = A_\mathrm{R}$. 
While both continuous and conventional \ac{MIMO} systems benefit from larger aperture sizes, the improvement in the conventional case grows more slowly. 
This occurs because increasing the aperture size in the conventional case only adds antennas in a discretized manner, which samples the aperture coarsely and loses some effective area. In contrast, the continuous \ac{MIMO} architecture scales more efficiently with aperture size, as the entire surface contributes electromagnetically through continuous current distributions.
Notably, between $0.25~\mathrm{m}^2$ and $0.30~\mathrm{m}^2$, no improvement is observed in the conventional case since the aperture increase does not add antennas under the discretization.
To verify whether the addition of more data streams limits performance gains, Fig.~\ref{fig:M_vary} shows results for varying numbers of \ac{RF} chains/data streams.
As seen, the received power remains unchanged, as the total transmit and receive power is evenly distributed across streams.

\begin{figure}[t!]
\centering
\includegraphics[width=\columnwidth]{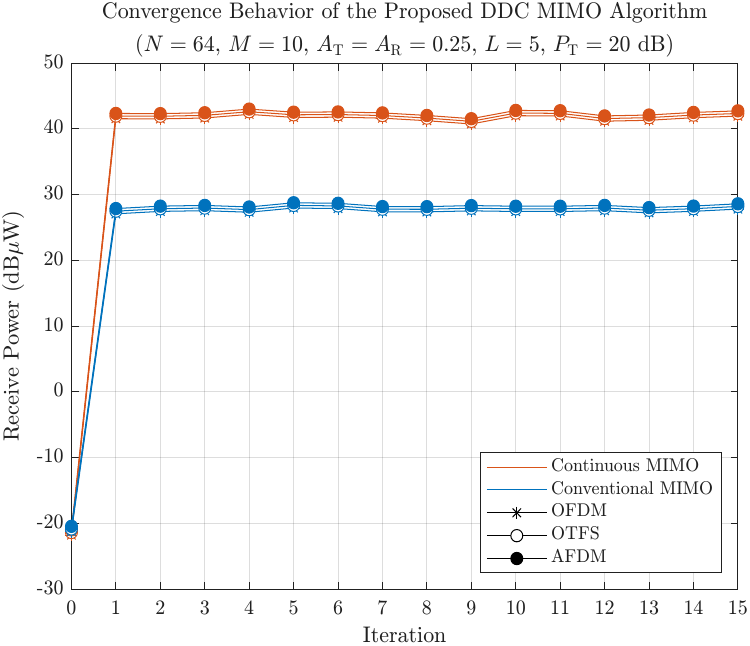}
\caption{Receive power vs. number of algorithm iterations for the DDC MIMO setup for \ac{OFDM}, \ac{OTFS} and \ac{AFDM}.}
\label{fig:convergence}
\end{figure}

Finally, Fig.~\ref{fig:convergence} illustrates the numerical convergence of the proposed Algorithm~\ref{alg:proposed_decoder}. 
The results confirm that the algorithm converges within only a few iterations, consistent with the convergence guarantees provided by the \ac{CoV}-based formulation.

\section{Conclusion}

We introduced a novel \ac{DD} channel model that integrates \ac{MIMO} \acp{CAPA} at both the \ac{TX} and \ac{RX}, referred to as the \ac{DDC MIMO} channel model. 
Within this framework, we derived explicit \ac{I/O} relationships for waveforms known to be well suited to \ac{DD} channels, namely \ac{OFDM}, \ac{OTFS}, and \ac{AFDM}. 
We further formulated and solved a functional optimization problem aimed at maximizing the received power using the \ac{CoV}, leading to closed-form solutions for both transmit and receive beamformers. 
These solutions resemble the classical matched filter expressions of conventional \ac{MIMO} systems, while naturally extending them to the continuous-aperture setting. 
Through simulations, we demonstrated that employing \ac{CAPA}-based architectures not only enables highly accurate modeling of massive \ac{MIMO} deployments, but also provides significant improvements in performance and computational efficiency. 
These results highlight the potential of continuous-aperture \ac{MIMO} to serve as a practical and scalable foundation for next-generation wireless networks, particularly in high-mobility environments characterized by \ac{DD} effects.

%\newpage
\bibliographystyle{IEEEtran}
\bibliography{references}

\end{document}